\documentclass[12pt,english,preprint]{aastex}
\usepackage{mathptmx}
\usepackage{helvet}
\usepackage{courier}
\usepackage[T1]{fontenc}
\usepackage[latin9]{inputenc}
\usepackage{graphicx}

\makeatletter
\usepackage{graphicx}

\def\la{\mathrel{\mathchoice {\vcenter{\offinterlineskip\halign{\hfil
\(\displaystyle##\)\hfil\cr<\cr\sim\cr}}}
{\vcenter{\offinterlineskip\halign{\hfil\(\textstyle##\)\hfil\cr
<\cr\sim\cr}}}
{\vcenter{\offinterlineskip\halign{\hfil\(\scriptstyle##\)\hfil\cr
<\cr\sim\cr}}}
{\vcenter{\offinterlineskip\halign{\hfil\(\scriptscriptstyle##\)\hfil\cr
<\cr\sim\cr}}}}}
\def\ga{\mathrel{\mathchoice {\vcenter{\offinterlineskip\halign{\hfil
\(\displaystyle##\)\hfil\cr>\cr\sim\cr}}}
{\vcenter{\offinterlineskip\halign{\hfil\(\textstyle##\)\hfil\cr
>\cr\sim\cr}}}
{\vcenter{\offinterlineskip\halign{\hfil\(\scriptstyle##\)\hfil\cr
>\cr\sim\cr}}}
{\vcenter{\offinterlineskip\halign{\hfil\(\scriptscriptstyle##\)\hfil\cr
>\cr\sim\cr}}}}}

\def\eg{{ e.g.,\ }}

\def\ie{{i.e.,\ }}

\def\ln{{\rm ln}}

\usepackage{hyperref}
\usepackage{esint}
\doublespace

\AtBeginDocument{
  
}

\makeatother

\usepackage{babel}

\begin{document}

\title{Probing Nearby CR Accelerators and ISM Turbulence with Milagro Hot
Spots}

\author{M. A. Malkov and P.H. Diamond}

\affil{CASS and Department of Physics, University of California, San Diego,
La Jolla, CA 92093-0424}

\author{L.O'C. Drury}

\affil{Dublin Institute for Advanced Studies, 31 Fitzwilliam Place, Dublin
2, Ireland}

\author{R.Z. Sagdeev}

\affil{University of Maryland, College Park, Maryland 20742-3280, USA}
\begin{abstract}
Both the acceleration of cosmic rays (CR) in supernova remnant shocks
and their subsequent propagation through the random magnetic field
of the Galaxy deem to result in an almost isotropic CR spectrum. Yet
the MILAGRO TeV observatory discovered a sharp ($\sim10^{\circ})$
arrival anisotropy of CR nuclei. We suggest a mechanism for producing
a weak and narrow CR beam which operates en route to the observer.
The key assumption is that CRs are scattered by a strongly anisotropic
Alfven wave spectrum formed by the turbulent cascade across the local
field direction. The strongest pitch-angle scattering occurs for particles
moving almost precisely along the field line. Partly because this
direction is also the direction of minimum of the large scale CR angular
distribution, the enhanced scattering results in a weak but narrow
particle excess. The width, the fractional excess and the maximum
momentum of the beam are calculated from a systematic transport theory
depending on a single scale $l$ which can be associated with the
longest Alfven wave, efficiently scattering the beam. The best match
to all the three characteristics of the beam is achieved at $l\sim1$pc.
The distance to a possible source of the beam is estimated to be within
a few 100pc. Possible approaches to determination of the scale $l$
from the characteristics of the source are discussed. Alternative
scenarios of drawing the beam from the galactic CR background are
considered. The beam related large scale anisotropic CR component
is found to be energy independent which is also consistent with the
observations.
\end{abstract}

\section{Introduction\emph{ }}

The MILAGRO TeV observatory recently discovered collimated beams dominated
by hadronic cosmic rays (CR) with a narrow ($\sim10^{\circ}$) angular
distribution in the 10 TeV energy range \citep{Milagro08PRL}. This
is surprising, since most of the CR acceleration and propagation models
predict only a weak, large scale anisotropy. The acceleration models
are based on the diffusive shock acceleration (DSA) mechanism, widely
believed to generate galactic CRs in supernova remnant shocks (SNR).
The corner stone of the DSA is a rapid pitch-angle scattering of CRs
by self-generated Alfven waves in the shock vicinity. An enhanced
scattering isotropizes particle distributions. Moreover, when the
shock releases the accelerated particles into the interstellar medium
(ISM), they continue to scatter by the ISM turbulence. Even though
this scattering occurs at a significantly lower rate, all sharp anisotropies
carried over from the accelerator or created otherwise, should be
erased during the long ($\ga$100 pc) travel of the CRs from any hypothetical
nearby SNR to the observer. Yet, the astounding sharp beaming effect
is argued to be genuine.

Focusing on relatively distant accelerators (such as nearby SNRs)
and long-distance propagation effects as a possible cause of the MILAGRO
beam(s), we do not consider 'local' scenarios that have already been
discussed and largely rejected by \citet{DruryAharMILAGRO08} and
\citet{SalvatiMilagro08}. As for the remote accelerator with subsequent
propagation effects, some of them have also been suggested in the
above publications. In particular, \citet{SalvatiMilagro08} associate
the observed CR beam with the Geminga pulsar. However, \citet{DruryAharMILAGRO08}
argue that this does not explain namely the sharp collimation, and
suggest a magnetic nozzle as such a collimation device. The magnetic
nozzle scenario, however, poses a rather strong constraint on the
nozzle mirror ratio ($B_{max}/B_{min}\sim\vartheta^{-2}\gg1,$ where
$\vartheta$ is the beam angular width). The advantage of this scenario
is that the beam density is equal to the difference between the isotropic
components of CRs on each side of the mirror (by linearity of the
transport equation). Since the Milagro beam is very weak ($\sim10^{-4}$
of the CR density), this is a very mild requirement on the CR enhancement
on the far side of the mirror. It is also true that the existence
of a magnetic mirror of that strength cannot be warranted or denied
on rational ground. It should be noted that any anisotropic distribution
may become vulnerable to self-spreading in pitch angle. As pointed
out by \citet{DruryAharMILAGRO08}, however, the isotropic CR background
should produce a stabilizing effect against the beam self-spreading.
We will quantify the CR stabilization in Sec.\ref{sec:BeamSustainability}
required for both the collimation mechanism suggested in the present
paper and for the magnetic nozzle hypothesis.

In this paper we suggest a novel mechanism for producing a narrow
CR beam. It is based on the strong anisotropy of the MHD turbulence
in the ISM. Such anisotropy is expected when the turbulence is driven
at a long (outer) scale, but unlike the isotropic Kolmogorov cascade,
the incompressible MHD cascade is directed perpendicularly to the
magnetic field in the wave vector space. This was shown by \citet{Goldr95}
(GS) (see also \citealt{SridhGoldr94} and \citealt{goldr97}) and
confirmed by numerical simulations (\eg \citealt{ChoVishn00,MaronGoldr01,BereznLaz09}).
The cascade proceeds to $k_{\perp}r_{g}\left(p\right)\gg1$ in the
perpendicular wave number direction for the protons with the gyro-radii
$r_{g}\sim10^{16}$cm, typical for the particles of the MILAGRO beam
energies $pc\sim10$TeV and the ISM magnetic field of a few $\mu G$.
Contrary to the $k_{\perp}$ direction the spectrum spreading in $k_{\parallel}$
is suppressed, so that $k_{\parallel}\sim k_{\perp}^{2/3}l^{-1/3}\ll k_{\perp}$,
where $l$ is the outer scale. 

\begin{figure}[h]
\includegraphics[bb=0bp 500bp 500bp 700bp]{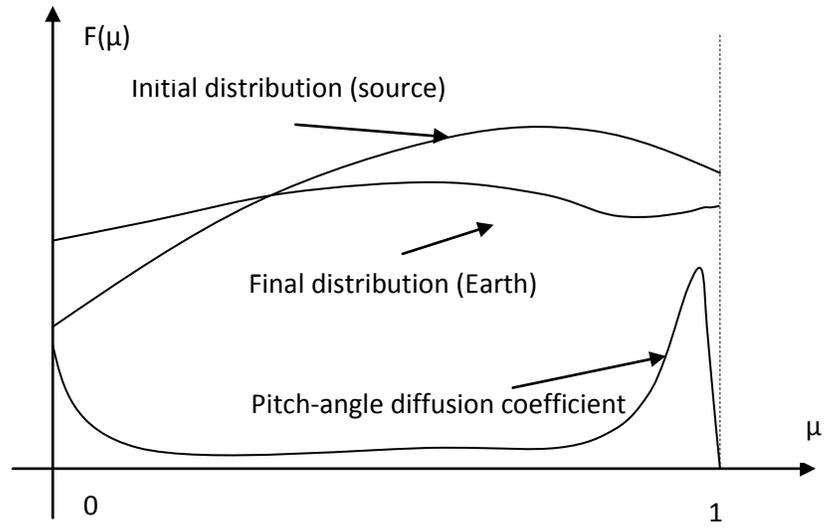}

\caption{Schematic representation of initial and final pitch-angle distributions
and that of the diffusion coefficient $D_{\mu\mu}\left(\mu\right)$.\label{fig:Pitch-angle-dependence}}

\end{figure}

As is known from the wave-particle interaction in plasmas, the scattering
of particles with the Larmor radius exceeding the wave length in the
perpendicular direction, $k_{\perp}r_{g}\gg1$ is strongly suppressed,
since such particles suffer a rapidly changing electromagnetic force.
Specifically, the CR scattering by the GS anisotropic spectrum was
investigated in a number of publications (\citealt{ChandranGS00PhRvL,YanLazar02}).
What is important for the purposes of this paper is that the pitch-angle
scattering rate is peaked at $\left|\mu\right|=\left|\cos\vartheta\right|\approx1$,
\ie for particles moving along the field line, since for these particles
$k_{\perp}r_{g}\left(p_{\perp}\right)\la1$. Only particles with such
small $p_{\perp}$, \ie with pitch angles within $\sin^{2}\vartheta\la\epsilon\ll1$
are scattered efficiently. Looking at this problem mathematically,
a peaked diffusion coefficient $D\left(\mu\right)$ does not necessarily
result in a peaked particle distribution $f\left(\mu\right)$. Indeed,
the time-asymptotic solution of the diffusion equation with zero flux
through the boundary is a flat distribution even if the diffusion
coefficient is not constant. Nevertheless, consider the particle diffusion
in pitch angle on an\emph{ intermediate time-scale}, \ie  when anisotropy
is erased within the strong peak of the diffusion coefficient $D\left(\mu\right)$,
but is present in the region where $D\left(\mu\right)$ is much smaller.
The dominant eigenfunction of the scattering operator has a relatively
broad minimum at $\mu=1$, \ie where $D$ is sharply peaked. Now,
the enhanced scattering fills up the very bottom of this minimum which
appears as a narrow excess, Fig.\ref{fig:Pitch-angle-dependence}.
In the context of a classical Lorentz gas relaxation problem \citep{Gurevich61RunAway,KruskalBernstein64},
this is clearly a transient effect associated with an incomplete decay
of the anisotropic part of the pitch-angle distribution. Note that
the difference with the Lorenz gas problem is in the sharply peaked
$D\left(\mu\right)$. In addition, our problem is a problem in $z$,
which is the spatial coordinate (rather than time) along a magnetic
flux tube that connects the CR source with the Earth. 

The demonstration of this phenomenon, facilitated and obscured at
the same time by the fact that the peak region $\left|\sin\vartheta\right|\ll1$
contains the singular points of particle pitch-angle diffusion operator
at $\sin\vartheta=0$, will be the main subject of the present paper.

Before we tackle this problem, we briefly discuss in the next section
how narrow the CR angular distribution can be as it leaks from a hypothetical
nearby SNR accelerator, magnetically connected with the heliosphere.
This, or some other moderately anisotropic distribution of CRs, created
by a recent SNR explosion, will be subjected in Sec.\ref{sec:Pitch-angle-scattering-of}
to the pitch-angle scattering analysis and to the propagation analysis
in Sec.\ref{sec:Particle-propagation}. Next, in Sec.\ref{sec:BeamSustainability}
we determine the maximum energy of the beam beyond which it must spread
on self-generated Alfven waves. Sec.\ref{Sec:Distance-to-the} deals
with the relation between the beam maximum energy and the distance
to its possible source. We conclude with a brief discussion of the
results and of what the fascinating MILAGRO findings can possibly
tell about a nearby accelerator and the structure of ISM turbulence.

\section{Angular distribution of Diffusively accelerated particles\label{sec:Angular-distribution-of}}

To estimate anisotropy of CRs escaping from a SNR accelerator, we
first briefly review the DSA mechanism and its possible modifications
that can enhance the CR anisotropy. Within this mechanism particles
gain energy by scattering upstream of the shock and back downstream
repeatedly. The scattering is supported by strong MHD waves unstably
driven by the accelerated particles themselves. In the early phase
of acceleration an ion-cyclotron instability dominates. It is driven
by a weak pitch-angle anisotropy of particle distribution. It is reasonable
to assume, however, that a small fraction of particles that reach
sufficiently high energies can diffuse to the distant part of the
turbulent shock precursor where their self-confinement becomes inefficient.
In this way a somewhat artificial notion of the 'free escape boundary'
(FEB) was introduced, particularly in Monte Carlo numerical schemes
\citep{Ellison1996} and other analytical and numerical studies \citep{BlasiEscape09,RevilleEsc09}.
Particle escape also occurs naturally if the plasma upstream is not
fully ionized and the weak wave excitation at the periphery of the
shock turbulent precursor is suppressed by the ion-neutral collisions
\citep{DruryNeutral96}. However, the angular distribution of particles
leaking through the FEB has not been calculated systematically. Note
that such calculation would require a self-consistent treatment of
wave generation and the relaxation of the distribution of leaking
particles. If the DSA process inside the precursor maintains CR isotropy,
the leaking particles may be assumed to have a one-sided quasi-isotropic
distribution. 

As the pressure of accelerated particles grows, other instabilities
may set on, including the non-resonant fire-hose instability \citep{Acht83FireHose,ShapiroQuest98GeoRL,Bell04}
and an acoustic instability driven by the \emph{pressure gradient}
upstream \citep{DruryFal86,ZankAM90,KangJR92}. From this point on,
the particle transport becomes more complicated. In particular, acoustic
waves turn into shocks and form a shock train which compresses magnetic
field and creates a scattering environment markedly different from
the weakly turbulent scattering field described above. It consists
of a random sequence of relatively weak shocks and was shown to produce
a loss cone in momentum space. However, preliminary calculations of
particle dynamics in this environment \citep{MD06} show that the
opening angle of run-away particles is still too large to account
for the MILAGRO observations, particularly when the subsequent self-spreading
of the beam is taken into account. This is clearly necessary since
the stabilization on the background CRs is not sufficient at this
phase of the beam propagation due to its relative strength. Apart
from the magnetic nozzle \citep{DruryAharMILAGRO08}, a remaining
option is to generate the beam on its way to the Earth.

At the first sight, this task appears to be like 'squeezing blood
out of a stone'. Intuitively, an intervening turbulence on the way
to the Earth, if anything, can only further spread the beam. That
the turbulent particle beaming is possible nonetheless, is primarily
due the very sharp dependence of the scattering frequency on the pitch
angle near the magnetic field direction.

\section{Pitch-angle scattering of CRs by anisotropic Alfven turbulence\label{sec:Pitch-angle-scattering-of}}

Systematic studies of the wave-particle interactions in magnetized
plasmas begun in early 60-s by \citealt{SagdShafr61,VVSQL62,RowlandsShapShev66}
and independently within the astrophysical and geophysical contexts
by \citealp{Jokipii66,Kennel66,Voelk73}. The angular profile of the
scattering frequency depends on the structure of turbulence. We provide
a concise derivation for the case of our interest in Appendix \ref{sec:AppendixA}.
More generally, the particle scattering by an anisotropic turbulence
with the spectrum suggested by \citealp{Goldr95} (GS) was studied
by \citet{ChandranGS00PhRvL}. He particularly demonstrated that the
maximum contribution to the pitch-angle scattering of the field aligned
particles is strongly dominated by the Alfven wave magnetic perturbations,
so that we neglect the contributions of magnetosonic waves and velocity
perturbations in what follows. The neglected components are essential
for particles with $\left|\mu\right|\ll1$, but we are primarily interested
in those with $\mu\approx1$, as they are assumed to make one of the
MILAGRO {}``hot spots''. \citet{ChandranGS00PhRvL} also gives a
detailed description of the pitch-angle diffusion coefficient for
the GS spectrum and identifies its peaks at $\left|\mu\right|=0,1$.
However, for the purposes of this paper we need the \emph{angular
profile} of the peak at $\left|\mu\right|=1$, which we evaluate in
the present section.

We begin with the general expression for the pitch-angle scattering
coefficient (\eg \citealp{Voelk73,ChandranGS00PhRvL}, Appendix \ref{sec:AppendixA}):

\begin{eqnarray}
D_{\mu\mu} & = & \Omega^{2}\left(1-\mu^{2}\right)\sum_{\mathbf{k},n}\frac{n^{2}J_{n}^{2}\left(\xi\right)}{\xi^{2}}\nonumber \\
 & \times & \intop_{0}^{\infty}I\left(k_{\parallel},k_{\perp},\tau\right)e^{i\left(k_{\parallel}v_{\parallel}+n\Omega\right)\tau}d\tau\label{eq:Dmm1}\end{eqnarray}
where we have used (standard) notations, provided in Appendix \ref{sec:AppendixA}.
Assuming the GS spectrum for the spectral wave density $I$ we have

\begin{equation}
I=\frac{1}{6\pi}k_{\perp}^{-10/3}l^{-1/3}g\left(\frac{k_{\parallel}l^{1/3}}{k_{\perp}^{2/3}}\right)e^{-\tau/\tau_{k}}\label{eq:I}\end{equation}
where we have assumed the notations and normalization of the spectrum
used by \citet{ChandranGS00PhRvL} rather than by GS. In particular
$g\left(x\right)=H\left(1-\left|x\right|\right)$, where $H$ is the
Heaviside function and $\tau_{k}=\left(l/V_{A}\right)\left(k_{\perp}l\right)^{-2/3}$
is the turbulence correlation time. Focusing on the resonant interactions
with particles, from eq.(\ref{eq:Dmm1}) we obtain

\begin{eqnarray}
D_{\mu\mu} & = & \frac{\pi}{3}l^{-1/3}\Omega^{2}\left(1-\mu^{2}\right)\intop_{0}^{\infty}k_{\perp}^{-7/3}dk_{\perp}\sum_{n=-\infty}^{\infty}\nonumber \\
 & \times & \frac{n^{2}J_{n}^{2}\left(\xi\right)}{\xi^{2}}\intop_{-\infty}^{\infty}g\left(\frac{k_{\parallel}l^{1/3}}{k_{\perp}^{2/3}}\right)\delta\left(k_{\parallel}v_{\parallel}-n\Omega\right)dk_{\parallel}\label{eq:Dmm2}\end{eqnarray}

Note that the integral in $k_{\perp}$ cuts off at the lower limit
by virtue of the function $g$. Therefore, from the last expression
we can get

\begin{equation}
D_{\mu\mu}=\frac{2\pi}{3}\frac{v}{l\left|\mu\right|}\left(1-\mu^{2}\right)y^{4/3}S\left(y\right)\label{eq:Dmm3}\end{equation}
where we have introduced the notation

\begin{equation}
S\left(y\right)=\sum_{n=1}^{\infty}S_{n}=\sum_{n=1}^{\infty}n^{2}\intop_{y\left(n/\left|\mu\right|\right)^{3/2}}^{\infty}J_{n}^{2}\left(x\right)x^{-q}dx\label{eq:S}\end{equation}
with $y=\sqrt{\left(1-\mu^{2}\right)/\epsilon}$, $\epsilon=v/l\Omega$
and $q=13/3$. Assuming $y>1$, we can take an asymptotic limit $x\gg1$
for $J_{n}$ and recover the corresponding result of \citet{ChandranGS00PhRvL}

\begin{equation}
D_{\mu\mu}\simeq\frac{2}{13}\zeta\left(\frac{9}{2}\right)\frac{v}{l}\epsilon^{3/2}\frac{\left|\mu\right|^{11/2}}{\sqrt{1-\mu^{2}}}\label{eq:Dmm4}\end{equation}
where $\zeta\left(s\right)=\sum_{n=1}^{\infty}n^{-s}$ is the Riemann
$\zeta-$ function. Note that $\zeta\left(9/2\right)\approx1.05$,
so that with a $5\%$ accuracy the $n=1$ term in eq.(\ref{eq:S})
would suffice. 

For larger values of $y$, namely when $\delta ln\left(1/\epsilon\right)\ga\epsilon^{3/2}\left|\mu\right|^{11/2}\left(1-\mu^{2}\right)^{-3/2}$,
where $\delta=V_{A}/v\approx V_{A}/c$, the finite correlation time
in the general form of $D_{\mu\mu}$ given by eqs.(\ref{eq:Dmm2}-\ref{eq:I})
should be taken into account. It is convenient to perform the integral
in $k_{\parallel}$ first, then perform that in $\tau$, which yields

\begin{eqnarray}
D_{\mu\mu} & = & \frac{1}{3}l^{-1/3}\frac{\Omega}{v\mu}\left(1-\mu^{2}\right)\sum_{n=-\infty}^{\infty}\nonumber \\
 & \times & \intop_{1/l^{\prime}}^{\infty}\frac{dk_{\perp}}{k_{\perp}^{7/3}}\tan^{-1}\left\{ \frac{1}{\delta}\left[\mu+\frac{n}{\epsilon}\left(k_{\perp}l\right)^{-2/3}\right]\right\} \frac{n^{2}}{\xi^{2}}J_{n}^{2}\left(\xi\right).\label{eq:Dmm5}\end{eqnarray}
In contrast to the previous case, the integral here needs to be cut
at the lower limit, by introducing the longest scale $l^{\prime}<l$
\citep{ChandranGS00PhRvL}. Perturbations with $k_{\perp}l^{\prime}\ga1$
scatter particles efficiently, while longer waves interact with particles
adiabatically. However, to simplify notations we set $l^{\prime}=l$
below, which is partly justified by a weak dependence of the turbulence
intensity on $l$, eq.(\ref{eq:I}). We will discuss our choice of
scales $l$ and $l^{\prime}$ in Sec.\ref{Sec:Distance-to-the}.

Expanding $\tan^{-1}$ for a large argument ($\delta,\epsilon\ll1$)
and summing the series of Bessel functions we obtain

\begin{eqnarray}
D_{\mu\mu} & \simeq & \frac{\delta\epsilon}{3}\Omega\left(1-\mu^{2}\right)\intop_{\epsilon\left(1-\mu^{2}\right)^{1/2}}^{\infty}\frac{d\xi}{\xi}\frac{1-J_{0}^{2}\left(\xi\right)}{\xi^{2}}\nonumber \\
 & \simeq & \frac{1}{6}\frac{v}{l}\delta\left[\ln\left(\frac{1}{\epsilon}\right)-\frac{1}{2}\ln\left(1-\mu^{2}\right)\right]\left(1-\mu^{2}\right)\label{eq:Dmm6}\end{eqnarray}
Again, within the assumed accuracy the complete sum with the Bessel
functions in eq.(\ref{eq:Dmm5}) yields approximately the same result
as only the terms with $n=\pm1$. We will show below that in calculating
the form of the peak of $D_{\mu\mu}\left(\mu\right)$ at $\left|\mu\right|\approx1$
it is sufficient to take only a few first terms into account. 

Now that we have reviewed the overall behavior of the pitch-angle
scattering frequency, we concentrate on the particular region, $1-\mu^{2}\la\epsilon$.
For, we evaluate the series in eq.$\left(\ref{eq:S}\right)$ for $y\ll1$.
It is clear that the main contributions comes from $n=1$, but since
we need also the sum for $y\sim1$, we should include a few next terms
and examine whether it will change the result substantially. Based
on the above remarks about the dominant contribution of the low $n$
terms, it will hopefully not. First we evaluate $S_{1}$ by integrating
it by parts and rearranging the remaining integrals as follows

\begin{eqnarray}
S_{1} & = & \frac{y^{1-q}}{q-3}J_{1}^{2}\left(y\right)-\frac{2}{q-3}\nonumber \\
 & \times & \left(\intop_{0}^{\infty}J_{1}\left(x\right)J_{2}\left(x\right)x^{1-q}dx-\intop_{0}^{y}J_{1}\left(x\right)J_{2}\left(x\right)x^{1-q}dx\right)\label{eq:S1}\end{eqnarray}
Note that the first term diverges as $y\to0,$ the second is finite
and the third one is small. Neglecting the third term, we obtain

\[
S_{1}\simeq\frac{3}{4}J_{1}^{2}\left(y\right)y^{-10/3}+S_{1}^{\prime}\]
where 

\[
S_{1}^{\prime}=-\frac{567}{6400\cdot2^{1/3}}\frac{\Gamma^{2}\left(1/3\right)}{\Gamma^{3}\left(2/3\right)}\simeq-.22\]
Adding to $S_{1}$ the leading in $y\ll1$ terms (constants) from
a few first $S_{n}$, and substituting thus obtained $S\left(y\right)$
into eq.(\ref{eq:Dmm3}) we arrive at the following final expression
for the scattering coefficient

\begin{equation}
D_{\mu\mu}=\frac{\pi}{2}\frac{v}{l}\left(1-\mu^{2}\right)\left[\frac{J_{1}^{2}\left(y\right)}{y^{2}}+ry^{4/3}\right]\label{eq:Dmmfin}\end{equation}
where $r\sim10^{-2}$ and $y=\sqrt{\left(1-\mu^{2}\right)/\epsilon}$.
Clearly, we can neglect the small second term in the brackets altogether,
and switch to the expression given by eq.(\ref{eq:Dmm6}) for $y\ga j_{1}$,
where $j_{1}\approx3.8$ being the first root of $J_{1}$. Summarizing
this section, the most important part of the scattering coefficient
$D_{\mu\mu}\left(y\right)$ is its sharp peak near $\left|\mu\right|=1$
where it behaves as $D_{\mu\mu}\propto J_{1}^{2}\left(y\right)$.
As $y$ grows and approaches $y=j_{1}$, $D_{\mu\mu}/\left(1-\mu^{2}\right)$
falls down to $\sim\delta$ of its peak value at $\left|\mu\right|=1$
and remains approximately constant, eq.(\ref{eq:Dmm6}). The other
peak occurs at $\mu\approx0$ but it is not important for our purposes.

\section{Particle propagation\label{sec:Particle-propagation}}

Suppose that a source of CRs is within the same magnetic flux tube
with the Earth. This source could either be a SNR currently accelerating
CRs which gradually escape from the accelerator or it could be due
to the CRs that have been accelerated not long ago, or any other region
of enhanced CRs. We calculate their propagation to the Earth below.
Obviously, the degree of CR anisotropy near the source may be significantly
higher than that observed at the Earth. The propagation problem may
be considered being one dimensional and stationary with the only spatial
coordinate $z$, directed along the flux tube from the source to the
Earth. However, we bear in mind the finite radius of the flux tube
by choosing the most important MHD mode that will scatter particles.
In particular, out of the three major MHD modes we select the Alfven
wave (with a dispersion relation $\omega=k_{\parallel}V_{A}$) since
it has no off-axis group velocity component and strong damping as
opposed to the fast and slow MHD waves. Note that in a box- rather
than in a thin tube-geometry the other modes are also essential for
the particle scattering \citep{YanLazar02,BereznLazar10}. On the
other hand, for $\mu\approx1$ propagation, the shear-Alfven wave
is still the most important mode \citep{ChandranGS00PhRvL}.

As the CR particles are assumed to be scattered by Alfven waves, almost
frozen into the local fluid, the particle momentum is conserved and
the transport problem is in only two variables, the coordinate $z$
and the pitch angle $\vartheta$ (or $\mu\equiv\cos\vartheta$). The
characteristic ($\vartheta$-independent) pitch-angle scattering frequency
$\nu_{\vartheta}$ (typical for $\mu$ not too close to $\mu=0,\pm1$,
where the pitch-angle diffusion coefficient has sharp peaks) can be
deduced from the previous section by unifying eqs.(\ref{eq:Dmm4})
and (\ref{eq:Dmm6}) (and omitting some factors which are close to
unity):

\begin{equation}
\frac{D_{\mu\mu}}{1-\mu^{2}}\approx\nu_{\vartheta}\equiv\frac{v}{l}\left(\delta\ln\left(\frac{1}{\epsilon}\right)+\epsilon^{3/2}\right)/6\label{eq:nu}\end{equation}
The equation for the CR distribution thus reads

\begin{equation}
(u+\mu)\frac{\partial f}{\partial z}=\frac{\partial}{\partial\mu}\left(1-\mu^{2}\right)D\left(\mu\right)\frac{\partial f}{\partial\mu}\label{eq:transp1}\end{equation}
Here $u$ is the bulk flow (scattering centers) velocity along $z$
in units of the speed of light, $u\ll1$, $\mu=\cos\vartheta$. The
coordinate $z$ is normalized to the pitch-angle scattering length
$c/\nu_{\vartheta}\approx v/\nu_{\vartheta}$, so that $D\left(\mu\right)=\nu_{\vartheta}^{-1}D_{\mu\mu}/\left(1-\mu^{2}\right)$
being normalized to $\nu_{\vartheta}$, is close to unity except for
the narrow peaks.

Our purpose is to find a narrow feature (which may be a bump or a
hole) on the otherwise almost isotropic angular spectrum $f\left(\mu\right)$.
Clearly, this feature must be pinned to one of the peaks of $D\left(\mu\right)$.
This feature will be shown to be weak, so it can be considered independent
of the other possible features on $f\left(\mu\right)$ that would
be related to the remaining two peaks on the function $D\left(\mu\right)$;
in other words, we apply a perturbative approach. 

Let us consider the particle scattering problem given by eq.(\ref{eq:transp1})
in a half space $z\ge0$ and assume that at $z=0$ (source) the distribution
function is $f\left(0,\mu\right)=f_{0}\left(\mu\right)$. Note that
$f_{0}$ is not quite arbitrary since it also contains particles coming
to the source (\ie those with $\mu<0$). A similar problem occurs
in the DSA at relativistic shocks \citep{KirkSchn87,KirkDuffyRelS99}
and in the problem of ion injection into the DSA \citep{mv95}. It
is clear that if there are no particle sources at $z=\infty$, then
$f\left(\infty,\mu\right)=f_{\infty}=const$, apart from the dependence
of $f$ on the particle momentum as a parameter. It is convenient
to subtract $f_{\infty}$ from $f$:

\begin{equation}
\Psi\left(z,\mu,p\right)=f\left(z,\mu,p\right)-f_{\infty}\left(p\right)\label{eq:PsiDef}\end{equation}
so that the new function $\Psi$ satisfies the same equation (\ref{eq:transp1})
as $f$ and the following boundary conditions

\[
\Psi=\left\{ \begin{array}{cc}
\phi\left(\mu\right)=f_{0}\left(\mu\right)-f_{\infty}, & z=0\\
0, & z=\infty\end{array}\right.\]
It is natural to expand the solution into the series of eigenfunctions
$\Psi_{\lambda}$

\begin{equation}
\Psi=\sum_{\lambda}C_{\lambda}\Psi_{\lambda}\left(\mu\right)e^{-\lambda z}\label{eq:PsiExp}\end{equation}
to be found from the following spectral problem

\begin{equation}
\frac{d}{d\mu}\left(1-\mu^{2}\right)D\left(\mu\right)\frac{d\Psi_{\lambda}}{d\mu}+\lambda\left(u+\mu\right)\Psi_{\lambda}=0\label{eq:SpProbl1}\end{equation}
As is well known (\citealp{Richardson1918}, see also \citealp{KirkSchn87}),
there exists a complete set of the orthogonal eigenfunctions $\left\{ \Psi_{\lambda}\right\} _{\lambda_{i}=-\infty}^{\lambda_{i}=\infty}$
with the discrete spectrum $\lambda_{i}$ having no limiting points
other than at $\pm\infty$. Therefore, the expansion coefficients
$C_{\lambda}$ are

\begin{equation}
C_{\lambda}=\frac{1}{\left\Vert \Psi_{\lambda}\right\Vert ^{2}}\intop_{-1}^{1}\left(u+\mu\right)\Psi_{\lambda}\left(\mu\right)\phi\left(\mu\right)d\mu\label{eq:Clamda}\end{equation}
where $\left\Vert \Psi_{\lambda}\right\Vert $ denotes the norm of
$\Psi_{\lambda}$. Clearly, $\phi\left(\mu\right)$ must satisfy the
set of conditions $C_{\lambda}=0$ for all $\lambda\le$0. This reflects
the fact that $\phi$ is not an arbitrary boundary condition, as we
already noted. Nevertheless, since particles predominantly propagate
into positive $z$-direction (away from the source), it is reasonable
to assume that $\phi\left(\mu>0\right)$ is larger than $\phi\left(\mu<0\right)$,
\ie the source creates the anisotropy. As usual, if we consider the
formal solution given by eq.(\ref{eq:PsiExp}) at such a distance
$z$ where $\left(\lambda_{2}-\lambda_{1}\right)z\ga1$, with $\lambda_{1,2}$
being the first (smallest) positive eigenvalues, the solution will
be dominated by the first eigenfunction $\Psi_{\lambda_{1}}.$ We
know that the anisotropy at the Earth is very small ($\sim10^{-3}$)
and, assuming it being not so small at the source, we deduce that
$\lambda_{1}z\gg1$ so that the inequality $\left(\lambda_{2}-\lambda_{1}\right)z\gg1$
should satisfy as well and we can limit our treatment of the spectral
problem given by eq.(\ref{eq:SpProbl1}) to the determination of only
the first positive eigenvalue with the corresponding eigenfunction.
Nevertheless, we return to this point in Sec.\ref{Sec:Distance-to-the}.
We also note that since $u\ll1$, we can set $u=0$ as there is no
significant influence of the region $\left|\mu\right|\ll1$ where
eq.(\ref{eq:ZeroOrdExt}) has a turning point, whereas we are primarily
interested in the behavior of the solution near a singular point at
$\mu=1$. Although the function $D\left(\mu\right)$ has a strong
peak at $\mu\approx1$, this peak is very narrow ($\sim\epsilon$)
and, as we mentioned, a perturbation theory applies. We start with
the outer solution, \ie  with the solution outside of the peak area.

\subsection{Angular distribution outside of the peak of the pitch-angle diffusion
coefficient\label{sub:AngDistOutsideBeam} }

\begin{figure}[h]
\includegraphics[bb=0bp 0bp 702bp 450bp,clip,scale=0.6]{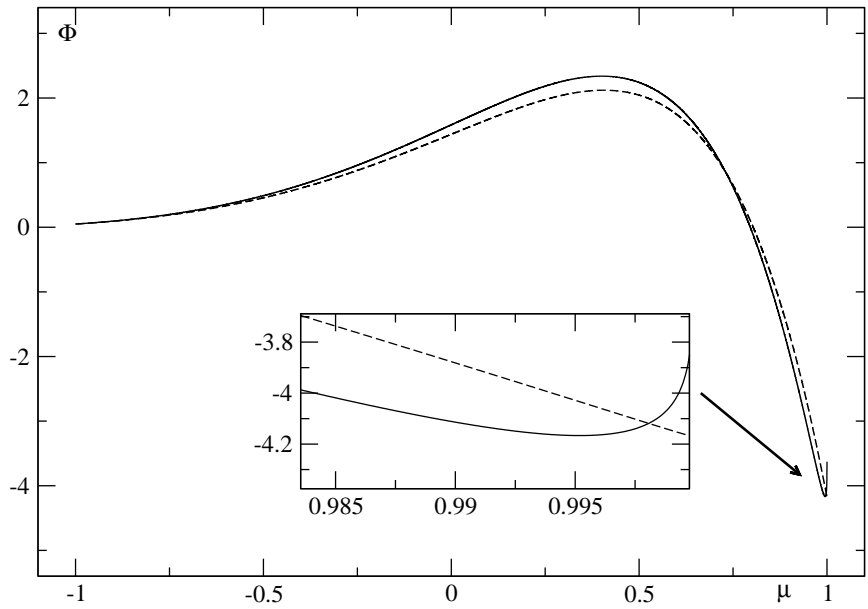}

\caption{Unperturbed eigenfunction $\Phi\left(\mu\right)\equiv\Psi_{\lambda_{1}}^{\left(0\right)}$
(numerical solution of eq.{[}\ref{eq:ZeroOrdExt}{]}, dashed line).
Perturbed solution (eqs.{[}\ref{eq:PsiPert}{]} and {[}\ref{eq:Psi1a}{]},
solid line). The insert shows the solution behavior at the end point,
including the logarithmic term of the outer solution. \label{fig:Eigenfunction}}

\end{figure}

Outside of the peak region (outer expansion) we assume $D=1$ as an
exact value for $D$. Therefore, for $\left(1-\mu^{2}\right)\ga\epsilon$,
the zeroth order approximation reads

\begin{equation}
\frac{d}{d\mu}\left(1-\mu^{2}\right)\frac{d\Psi_{\lambda}^{\left(0\right)}}{d\mu}+\lambda^{\left(0\right)}\mu\Psi_{\lambda}^{\left(0\right)}=0\label{eq:ZeroOrdExt}\end{equation}
To find $\lambda^{\left(0\right)}$ we require the solution to be
regular at the both singular points $\mu=\pm1$. The solution of this
problem can be found by a number of numerical methods, for example,
by decomposing $\Psi_{\lambda}^{\left(0\right)}$ in a series of Legendre
polynomials \citep[e.g.,][and references therein]{KirkSchn87}. Since
we have set $u=0$ (as opposed to the cited paper where $u\simeq1)$,
as few as the first six polynomials would suffice, with a cubic equation
for $\lambda$. However, eq.(\ref{eq:ZeroOrdExt}) contains no parameters
(except $\lambda$, of course) so that the most practical approach
is to find the required single eigenvalue $\lambda_{1}$ and the corresponding
eigenfunction by a direct numerical integration of the above equation.
The result is shown in Fig.\ref{fig:Eigenfunction} and $\lambda_{1}\approx14.54$.
Since $\lambda_{2}\simeq2\lambda_{1}$, the WKB approximation can
be applied for all $\lambda\ge\lambda_{2}$ points of the spectrum.
However, the first eigenfunction and the eigenvalue $\lambda_{1}$
is sufficient for our purposes. 

Since $D\equiv1$ in the outer region, the perturbation can be associated
only with the perturbation of $\lambda$. Therefore, we expand $\lambda$
and $\Psi_{\lambda}$ as

\begin{equation}
\lambda=\lambda^{\left(0\right)}+\delta\lambda+\dots\label{eq:LamdaPert}\end{equation}

\begin{equation}
\Psi_{\lambda}=\Psi_{\lambda}^{\left(0\right)}+\delta\lambda\Psi_{\lambda}^{\left(1\right)}+\dots\label{eq:PsiPert}\end{equation}
Here $\lambda$ can be an arbitrary point of the spectrum $\lambda=\lambda_{i}>0$,
but we are primarily interested in the case $\lambda=\lambda_{1}$.
The equation for $\Psi_{\lambda}^{\left(1\right)}$ takes the following
form 

\begin{equation}
\frac{d}{d\mu}\left(1-\mu^{2}\right)\frac{d\Psi_{\lambda}^{\left(1\right)}}{d\mu}+\lambda^{\left(0\right)}\mu\Psi_{\lambda}^{\left(1\right)}=-\mu\Psi_{\lambda}^{\left(0\right)}\label{eq:ZerOrdInhom}\end{equation}
Since the r.h.s. of this equation is not orthogonal to the solution
of its homogeneous part (eq.{[}\ref{eq:ZeroOrdExt}{]}), the operator
on the l.h.s. of eq.(\ref{eq:ZerOrdInhom}) is not identical to that
in eq.(\ref{eq:ZeroOrdExt}). Namely, the regularity condition at
$\left|\mu\right|=1$ no longer applies. Instead, a singular, linearly
independent counterpart of the solution of eq.(\ref{eq:ZeroOrdExt})
should be included (which is appropriate for the outer solution, but
which is not for the inner solution that will be considered in the
next subsection). Note that being interested in the behavior of the
overall solution near $\mu=1$, we can still require the solution
being regular at $\mu=-1$, since the unperturbed eigenfunction is
small there (\eg Fig. \ref{fig:Eigenfunction}) and the perturbation
at $\mu\simeq-1$ does not significantly influence the overall behavior
of the solution. With this in mind, we can write the solution of the
last equation as follows

\begin{equation}
\Psi_{\lambda}^{\left(1\right)}=-\Phi\intop_{-1}^{\mu}\frac{U\left(\mu^{\prime}\right)d\mu^{\prime}}{\Phi^{2}\left(\mu^{\prime}\right)\left(1-\mu^{\prime2}\right)}\label{eq:Psi1}\end{equation}
where we have denoted $\Phi\left(\mu\right)\equiv\Psi_{\lambda}^{\left(0\right)}\left(\mu\right)$,
and 

\begin{equation}
U\left(\mu\right)\equiv\intop_{-1}^{\mu}\mu^{\prime}\Phi^{2}\left(\mu^{\prime}\right)d\mu^{\prime},\label{eq:Udef}\end{equation}
for short. Two further remarks are in order here. First, the above
solution diverges logarithmically when $\mu\to1$. But, it is not
applicable within $1-\mu\la\epsilon$, where an inner expansion should
be obtained and matched to the solution, given by the outer expansion,
eqs.(\ref{eq:PsiPert}) and (\ref{eq:Psi1}). Second, the integral
in eq.(\ref{eq:Psi1}) is improper because $\Phi\left(\mu\right)$
has zeroes, in particular the one at $\mu=\mu_{1}\simeq0.8$ for $\lambda=\lambda_{1}$.
The integral should be understood in terms of a principle value and
the solution behaves at $\mu\approx\mu_{1}$ as $\Psi_{\lambda_{1}}^{\left(1\right)}\propto\left(\mu-\mu_{1}\right)\ln\left|\mu-\mu_{1}\right|$.
Before we turn to the inner part of the solution, for the purpose
of matching, it is convenient to rewrite the outer solution, given
by eq.(\ref{eq:Psi1}), in the following form

\begin{eqnarray}
\Psi_{\lambda}^{\left(1\right)} & = & \frac{U\left(1\right)}{2\Phi^{2}\left(1\right)}\Phi\left(\mu\right)\ln\left(\frac{1-\mu}{2}\right)-\Phi\left(\mu\right)\nonumber \\
 & \times & \intop_{-1}^{\mu}\frac{d\mu^{\prime}}{1-\mu^{\prime}}\left[\frac{U\left(\mu^{\prime}\right)}{\left(1+\mu^{\prime}\right)\Phi^{2}\left(\mu^{\prime}\right)}-\frac{U\left(1\right)}{2\Phi^{2}\left(1\right)}\right]\label{eq:Psi1a}\end{eqnarray}
Here the first term is singular at $\mu=1$ while the second term
is regular there.

\subsection{Angular distribution of the beam\label{sub:AngDistInsideBeam}}

Turning to the inner expansion of the solution of eq.(\ref{eq:SpProbl1}),
it is convenient to stretch the variable $\mu$ at $\mu=1$ as follows

\begin{equation}
w=\frac{1-\mu}{b}\label{eq:wvar}\end{equation}
Note that $b=\epsilon j_{1}^{2}/2$ is chosen in such a way that $D\left(w=1\right)\approx1$
(see Sec.\ref{sec:Pitch-angle-scattering-of}). Therefore, we represent
$D$ as 

\begin{equation}
D\left(w\right)=\left\{ \begin{array}{cc}
a^{-1}F\left(w\right)+1, & w\le1\\
1, & w>1\end{array}\right.\label{eq:Fdef}\end{equation}
with

\begin{equation}
F\left(w\right)=\frac{\pi}{2}\frac{1}{j_{1}^{2}w}J_{1}^{2}\left(j_{1}\sqrt{w}\right),\ \ \ w\le1\label{eq:Fdef1}\end{equation}
and $F\equiv0$ for $w>1$. Note that $a=\nu_{\vartheta}l/v\ll1$,
sec.\ref{sec:Pitch-angle-scattering-of}. Eq.(\ref{eq:SpProbl1})
can be written as follows

\begin{equation}
\frac{d}{dw}\left[F\left(w\right)+a\right]\left(2-bw\right)w\frac{d\Psi_{\lambda}^{i}}{dw}+ba\lambda\left(1-bw\right)\Psi_{\lambda}^{i}=0\label{eq:InnEq}\end{equation}
where the index $i$ stands for the 'inner' solution. In contrast
to the outer problem we must impose the regularity condition at $\mu=1$
($w=0$). Since $F$ vanishes for $w>1$ the expansion of the solution
of this equation should be sought for in a series of powers of $b$,
not $ba$, as an inspection of the second term of the equation may
suggest. The latter form of the expansion would be valid only for
$w<1$, whereas we need to match the solution of eq.(\ref{eq:InnEq})
with the outer solution at $w\ga1.$ While the regularity condition
at $w=0$ fixes one of the two arbitrary constants of the solution,
it is convenient to choose the second constant as the value of the
solution at $w=0$, \ie $\Psi_{\lambda}^{i}\left(0\right)$.

Working up to the second order in $b\ll1$, and integrating eq.(\ref{eq:InnEq})
by parts, we transform it into the following first order equation

\[
\frac{d\Psi_{\lambda}^{i}}{dw}+\frac{\lambda b}{2}g^{\prime}\left[1-\frac{\lambda b}{2}\left(\frac{h}{w}-g\right)\right]\Psi_{\lambda}^{i}=0\]
with the following obvious solution

\begin{equation}
\Psi_{\lambda}^{i}\left(w\right)=\Psi_{\lambda}^{i}\left(0\right)\exp\left\{ -\frac{\lambda b}{2}g\left(1+\frac{\lambda b}{4}g\right)+\frac{\lambda^{2}b^{2}}{4}\int_{0}^{w}g^{\prime}hdw/w\right\} \label{eq:InnExp1}\end{equation}
where we have used the notation

\begin{equation}
g\left(w\right)=a\intop_{0}^{w}\frac{dw^{\prime}}{F\left(w^{\prime}\right)+a}\label{eq:gdef}\end{equation}
and \[
h=\intop_{0}^{w}g\left(w^{\prime}\right)dw^{\prime}.\]
As it is seen from eq.(\ref{eq:Fdef}), the function $g$ can be evaluated
as follows 

\begin{eqnarray}
g & = & \left\{ \begin{array}{cc}
\frac{8}{\pi}aw, & w<1\\
g_{1}+w-1, & w\ge1\end{array}\right.\label{eq:gcases}\end{eqnarray}
where $g_{1}\simeq-\sqrt{2\pi a}/J_{0}\left(j_{1}\right)$. In order
to match the inner solution given by eq.(\ref{eq:InnExp1}) with the
outer solution obtained earlier (see eqs.{[}\ref{eq:PsiPert},\ref{eq:Psi1a}{]}),
we need to expand both solutions in power of $w$ or $1-\mu$ which
are valid in an overlap region. This is obviously a region where $1-\mu\ll1$
(to make a series expansion of the outer solution accurate) and $w>1$
(to make the inner solution simple, e.g., to use eq.{[}\ref{eq:gcases}{]}
for $g$). As $w=\left(1-\mu\right)/b$ with $b\ll1$, the overlap
region exists. Let us write the outer solution given by eqs.(\ref{eq:PsiPert},\ref{eq:Psi1a})
in terms of the inner variable 

\begin{eqnarray}
\Psi_{\lambda} & = & \Phi\left(1\right)-\Phi^{\prime}\left(1\right)bw+\frac{1}{2}\Phi^{\prime\prime}\left(1\right)b^{2}w^{2}\nonumber \\
 & + & \frac{U\left(1\right)}{2\Phi\left(1\right)}\delta\lambda\left(\ln w+\ln b\right)+\mathcal{O}\left(\delta\lambda+b^{3}\right)\label{eq:OutExpFin}\end{eqnarray}
where the primes denote the derivatives of $\Phi\left(\mu\right)$
at $\mu=1$. The first three terms of the last expression is nothing
but the leading terms of the Frobenius expansion of the unperturbed
regular part of the solution of eq.(\ref{eq:ZeroOrdExt}) at the singular
end point $\mu=1$. The fourth term is a perturbative, singular part
of expansion, which is entirely due to the fact that the spectral
parameter $\lambda$ deviates from its eigenvalue, ie $\delta\lambda=\lambda-\lambda_{i}\neq0$.
The both expansions are written in terms of the inner variable $w>1$,
and thus $D\left(\mu\right)=1$ in eq.(\ref{eq:SpProbl1}).

The inner solution given by eq.(\ref{eq:InnExp1}) can be written
for $w>1$ as follows

\begin{equation}
\Psi_{\lambda}^{i}=\Psi_{\lambda}^{i}\left(0\right)\left[1-\frac{1}{2}\lambda b\left(g_{1}+w-1\right)+\frac{1}{8}\lambda^{2}b^{2}\left(\frac{1}{2}w^{2}-2w+\ln w\right)\right]\label{eq:InExpFin}\end{equation}
Comparing the last two results, we deduce

\begin{equation}
\Psi_{\lambda}^{i}\left(0\right)\approx\frac{\Phi\left(1\right)}{1+\lambda b/2}\label{eq:InOutRel0}\end{equation}
and \begin{equation}
\delta\lambda=\frac{\lambda^{2}\Phi^{2}\left(1\right)}{4U\left(1\right)}b^{2}.\label{eq:DeltaLamda}\end{equation}
Using the matching procedure we determined the initially unknown arbitrary
constant of the inner solution $\Psi_{\lambda}^{i}\left(0\right)$
and the perturbation of the eigenvalue $\lambda$ by matching the
terms in both equations that are independent of $w$ and proportional
to $\ln w$, respectively. The linear and quadratic terms in $w$
match automatically to the appropriate accuracy $\sim b^{2}$. This
follows from the two further relations 

\[
\Phi^{\prime}\left(1\right)=\frac{\lambda}{2}\Phi\left(1\right),\;\;\;\;\Phi^{\prime\prime}\left(1\right)=\frac{\lambda}{4}\Phi^{\prime}\left(1\right),\]
which can be obtained from the Frobenius series of eq.(\ref{eq:ZeroOrdExt})
at the singular end point $\mu=1$ with $\Psi_{\lambda}^{(0)}\equiv\Phi$.

The following two observations are important for the goal of this
paper. First, since $U\left(1\right)$ in eq.(\ref{eq:Udef}) is positive
definite (which may be readily seen from the equation for $\Phi$,
eq.{[}\ref{eq:ZeroOrdExt}{]}, by multiplying it by $\Phi$ and integrating
by parts), $\delta\lambda$ in eq.(\ref{eq:DeltaLamda}) is positive
definite as well. Of course, this property of the spectrum can be
seen directly from eq.(\ref{eq:SpProbl1}) by virtue of the positive
sign of the perturbation of the coefficient $D$. Second, the perturbed
absolute value of the eigenfunction at $\mu=1$, given by eq.(\ref{eq:InOutRel0})
is always less than the unperturbed one, \ie $\left|\Psi_{\lambda}^{i}\left(0\right)\right|<\left|\Phi\left(1\right)\right|$.
Below, we discuss the observational consequences of these results.

\subsection{Observational Appearance of the Beam\label{sub:Observational-consequences}}

After we have determined the angular distribution of the beam, the
question is whether it is consistent with at least the prominent MILAGRO
hot spot A \citep{Milagro08PRL}. There are two observationally testable
properties of the solution. The first property is that the sign of
the perturbative correction to the distribution function is opposite,
according to eqs.(\ref{eq:InExpFin},\ref{eq:InOutRel0}), to the
sign of $\Phi\left(1\right)$. Of course, the latter can be changed
arbitrarily but then the main expansion coefficient, eqs.(\ref{eq:PsiExp},\ref{eq:Clamda})
will change its sign as well. Let us fix $\Phi\left(1\right)\equiv\Psi_{\lambda_{1}}^{\left(0\right)}\left(1\right)<0$,
as shown in Fig.\ref{fig:Eigenfunction}. Then, the perturbative correction
will produce a logarithmic peak at $\mu=1$ if $C_{\lambda_{1}}>0$
and a negative logarithmic hole in the opposite case. As it may be
seen from eq.(\ref{eq:Clamda}) and Fig.\ref{fig:Eigenfunction},
the requirement $C_{\lambda_{1}}>0$ constrains the initial distribution
$f_{0}\left(\mu\right)$ as follows. While being enhanced in the region
$\mu>0$ (particles propagate predominantly away from the source towards
the Earth), $f_{0}\left(\mu\right)$ should not be concentrated too
close to $\mu\approx1$. One simple conclusion from this observation
is that if an accelerator at $z=0$ produces a beam of leaking particles,
this beam should not be collimated tightly at $\mu=1$, or even overpopulate
the region $\mu_{1}<\mu<1$, where $\Phi\left(\mu_{1}\right)=0$ with
$\mu_{1}\simeq0.8$.

The second property of the solution is that the bump on the particle
angular distribution is located at the local minimum of the unperturbed
eigenfunction. This is formally not consistent with the MILAGRO results.
On the contrary, the latter indicate that the bump is in the region
of monotonic change of the distribution function. However, the particle
distribution obtained above is coming only from the flux tube that
connects the Earth with the source of the beam. Therefore, to obtain
the total distribution, the CR background anisotropic component should
be added. The latter, being independent of the beam source, is most
likely to change monotonically in any randomly selected area, such
as the MILAGRO hot spot A, so there is no apparent contradiction in
this regard.

In Sec. \ref{sub:AngDistInsideBeam} the two major beam parameters
were calculated in terms of the small parameter of the theory, 

\[
\epsilon=\frac{r_{g}\left(p\right)}{l}\]
where $r_{g}$ is the particle gyro-radius and $l$ is maximum wave
length beyond which particles interact with waves adiabatically. The
first parameter of the beam is its angular width (in terms of $\mu=\cos\vartheta$)

\begin{equation}
b=\frac{j_{1}^{2}}{2}\epsilon\approx7.3\epsilon\label{eq:bBeam}\end{equation}
and the second is its strength, which can be conveniently expressed
as the ratio of the beam excess to the amplitude of the first eigenfunction,
eq.(\ref{eq:InOutRel0})

\begin{equation}
\frac{\delta\Phi\left(1\right)}{\Phi\left(1\right)}\approx\frac{1}{2}\lambda_{1}b\approx53.4\epsilon\label{eq:BeamStrength}\end{equation}
Since $\epsilon\propto p$, the spectrum of the beam should be one
power harder than the CR large scale anisotropic component inside
the flux tube. This is consistent with the Milagro beam spectrum,
provided that $\Phi$ scales with momentum similarly to the galactic
CR background.

According to the MILAGRO Region A observations, the beam width is
about $\Delta\vartheta\sim10^{\circ}$, where $\Delta\vartheta\approx\cos^{-1}\left(1-b\right)\approx\sqrt{2b}=j_{1}\sqrt{\epsilon}$
so that we obtain for $\epsilon$ the following constraint from the
observed MILAGRO Spot A 

\[
\epsilon\approx\left(\frac{\Delta\vartheta}{j_{1}}\right)^{2}\approx2.1\cdot10^{-3}.\]
This estimate yields the strength of the beam given by eq.(\ref{eq:BeamStrength})
at the level of $\approx0.1$ which is also consistent with the MILAGRO
fractional excess of the beams A and B measured with respect to the
large scale anisotropy. 

In this section we made a preliminary consistency check of the beam,
as it forms while the large scale anisotropic CR distribution propagates
from its source to the Earth. In the next section we verify conditions
under which the beam can really reach the Earth without self-destruction,
as it is well known that beams in plasmas readily go unstable.

\subsection{Beam Sustainability\label{sec:BeamSustainability}}

Now that we have calculated the pitch-angle distribution of a narrow
CR beam formed from a wide angle anisotropic CR flux by its interaction
with the \emph{background ISM turbulence}, we need to check whether
the beam will survive the pitch-angle scattering by \emph{self-generated
waves}. The threat is the cyclotron instability of the beam but the
hope (as already mentioned by \citet{DruryAharMILAGRO08} in regard
with the magnetic nozzle) is the isotropic part of the CR background
distribution which should stabilize the beam. The dispersion relation
is a standard one, which can be written as follows (see \eg  \citealp{Acht83FireHose})

\begin{eqnarray}
1-\frac{\omega^{2}}{k^{2}V_{A}^{2}}+\frac{4\pi^{2}e^{2}}{k}\int\frac{p_{\perp}^{2}dp_{\perp}dp_{\parallel}}{p^{2}\left(kv_{\parallel}\pm\omega_{c}-\omega\right)}\nonumber \\
\times\left[p_{\perp}\frac{\partial F}{\partial p_{\parallel}}-\left(p_{\parallel}-\frac{\omega p}{kc}\right)\frac{\partial F}{\partial p_{\perp}}\right]=0\label{eq:DispEq}\end{eqnarray}
where $\pm$ signs correspond to the left/right polarized Alfven waves
propagating along the field line at the Alfven speed $V_{A}$ ($k\approx k_{\parallel},\,$$\omega\approx\pm kV_{A}$).
The distribution function $F\left(p_{\parallel},p_{\perp}\right)$
refers to the sum of the isotropic CR background distribution $F_{C}$,
the beam distribution $F_{B}$, and a large scale anisotropic part
$F_{1}\left(p_{\parallel},p_{\perp}\right)$. The latter, in turn,
consists of both the unperturbed solution $\Phi\left(\mu,p\right)$,
obtained in Sec.\ref{sub:AngDistOutsideBeam} and the background large
scale anisotropic component, most likely not related to the source
of the beam. Thus, the total distribution function can be represented
as $F=F_{C}\left(p\right)+F_{B}\left(p_{\parallel},p_{\perp}\right)+F_{1}\left(p_{\parallel},p_{\perp}\right)$.
Since the beam is concentrated at small pitch angles, \ie $0<p_{\perp}\ll p_{\parallel}$,
we assume its contribution to be larger than that of $F_{1}$. Clearly,
both $F_{C}$ and $F_{B}$ are small compared to the background plasma
density which yields the second term in eq.(\ref{eq:DispEq}). 

To simplify the calculations, it is convenient to introduce the new
variable $\rho$, instead of $p_{\perp}$ 

\[
\rho=p-\delta p_{\parallel}\equiv\sqrt{p_{\parallel}^{2}+p_{\perp}^{2}}-\delta p_{\parallel}\]
here $\delta=\pm V_{A}/c$, where $\pm$ relates to the forward and
backward propagating waves ($\omega=\pm kV_{A}$), respectively. Note
that the lines of constant $\rho$ coincide with the lines of quasilinear
diffusion of the distribution function $F$ and with the direction
of differentiation in the brackets in eq.(\ref{eq:DispEq}). On writing,
$\omega=\pm kV_{A}+\gamma$, ($\gamma\ll kV_{A}$) and neglecting
a small term $\delta\ll1$ in the resonance denominator of eq.(\ref{eq:DispEq}),
we obtain for the wave growth-rate the following relation

\[
\gamma=\frac{2\pi^{3}e^{2}}{\left|k\right|}\delta\int\frac{p_{\perp}^{3}}{p}\delta\left(p_{\parallel}\pm\frac{eB_{0}}{kc}\right)\left.\frac{\partial F}{\partial p_{\parallel}}\right|_{\rho}dp_{\perp}dp_{\parallel}.\]
Since $F_{C}=F_{C}\left(p\right)$ and $\left.\partial F_{C}/\partial p_{\parallel}\right|_{\rho}\approx\delta\partial F_{C}/\partial p$,
the contribution of $F_{C}$ into the growth-rate is stabilizing ($\partial F_{C}/\partial p<0$)
for both signs of $\delta$ and both wave polarizations. The contribution
of the beam is destabilizing, also for both polarizations, as long
as the real part of the frequency is taken as $\omega\approx+kV_{A}$,
\ie $\delta>0$. If the beam density were above the instability threshold,
it would rapidly spread in pitch angle on the self-generated waves
along the lines $\rho=const$. Therefore, the beam momentum distribution
can be obtained from the instability threshold condition, \ie from
an assumption that the imaginary contributions from the beam and from
the background CRs cancel. Thus, splitting the total distribution
as $F=F_{C}\left(p\right)+F_{B}\left(p_{\parallel},p_{\perp}\right)$
and integrating the term with $F_{B}$ by parts in $p_{\perp}$, we
obtain for the growth rate ($0<\delta\ll1$)

\begin{eqnarray}
\gamma & = & \frac{2\pi^{3}e^{2}}{\left|k\right|}\delta\intop_{0}^{\infty}\frac{p_{\perp}^{3}}{p^{2}}dp_{\perp}dp_{\parallel}\delta\left(p_{\parallel}-\frac{eB_{0}}{\left|k\right|c}\right)\nonumber \\
 & \times & \left(2\frac{p_{\parallel}^{2}}{p_{\perp}^{2}}F_{B}+\delta p\frac{\partial F_{c}}{\partial p}\right)\label{eq:GrowthRate}\end{eqnarray}
The beam contribution to the growth rate suggests to introduce a beam
distribution integrated in $p_{\perp}$: 

\begin{equation}
\mathcal{F}_{B}\left(p_{\parallel}\right)\equiv\frac{1}{p_{\parallel}^{2}}\int p_{\perp}F_{B}\left(p_{\perp},p_{\parallel}\right)dp_{\perp}\label{eq:InegrBeamDistr}\end{equation}
Note that for the beam particles $p_{\parallel}\approx p$. Assuming
a power-law momentum scaling for $F_{C}\left(p\right)\propto p^{-q_{c}}$
(with $q_{c}=4.6-4.7$, appropriate for the background CR momentum
distribution) from eq.(\ref{eq:GrowthRate}) we can obtain an expression
for the instability threshold distribution $\mathcal{F}_{th}\left(p_{\parallel}\right)$.
As we noted, this is the beam distribution that cancels $\gamma$
in eq.(\ref{eq:GrowthRate}):

\begin{equation}
\mathcal{F}_{th}\left(p_{\parallel}\right)\equiv\frac{\delta}{q_{c}-2}F_{C}\left(p_{\parallel}\right)\label{eq:Thresh}\end{equation}

so that if $\mathcal{F}_{B}\left(p_{\parallel}\right)\le\mathcal{F}_{th}\left(p_{\parallel}\right)$,
the beam can sustain its angular distribution. Otherwise, it will
be spread in pitch angle to satisfy the last inequality. Assuming,
however that this inequality holds, we calculate $\mathcal{F}_{B}$
using our results from the previous section. First, unlike the threshold
function $\mathcal{F}_{th}\left(p_{\parallel}\right)$ which is determined
by the isotropic CR background, the momentum dependence of $\mathcal{F}_{B}\left(p_{\parallel}\right)$
is prescribed by the wide angle anisotropic component, denoted earlier
as $\Phi\left(\mu\right)$, eqs.(\ref{eq:OutExpFin}-\ref{eq:InOutRel0})
and (\ref{eq:BeamStrength}). The particle momentum entered this function
as a parameter (which we omitted, for short) since we were considering
only the pitch-angle scattering under the conserved momentum. Using
the expressions for the width of the beam and for its amplitude relative
to $\Phi\left(\mu,p\right)$ given by eqs.(\ref{eq:bBeam}) and (\ref{eq:BeamStrength}),
respectively, we can represent $\mathcal{F}_{B}\left(p_{\parallel}\right)$
as follows

\begin{equation}
\mathcal{F}_{B}\left(p_{\parallel}\right)=\frac{\lambda_{1}b^{2}}{2}F_{0}\left(p_{\parallel}\right)=\frac{1}{8}\lambda_{1}j_{1}^{4}\epsilon^{2}F_{0}\left(p_{\parallel}\right)\label{eq:Fb2}\end{equation}
where we have denoted $F_{0}\left(p\right)\equiv\Phi\left(\mu=1,p\right)$.
Then, our constraint $\mathcal{F}_{B}\left(p_{\parallel}\right)\le\mathcal{F}_{th}\left(p_{\parallel}\right)$
can be represented in the following way

\begin{equation}
F_{0}\left(p\right)\le A\frac{V_{A}}{c}\frac{l^{2}}{r_{g}^{2}\left(p\right)}F_{C}\left(p\right)\label{eq:FinConstr}\end{equation}
where $r_{g}=pc/eB_{0}$ is the particle gyro-radius. We denoted by
$A$ the following numerical factor

\[
A=\frac{8}{\lambda_{1}j_{1}^{4}\left(q_{c}-2\right)}\approx10^{-3}\]
Due to the factor $r_{g}^{-2}$ in the relation given by eq.(\ref{eq:FinConstr}),
the function $F_{0}\left(p\right)$ is constrained at high momenta.
Assuming that $F_{0}$ is not much steeper than the background distribution
$F_{C}$, we infer from eq.(\ref{eq:FinConstr}) that there exists
maximum momentum $p_{Bmax}$, beyond which the beam would spread in
pitch-angle and dissolve in the CR background. In fact we can extract
more information from the last constraint. To conform with the Milagro
results, we assume $F_{0}\sim\tilde{F}_{C}$, where $\tilde{F}_{C}$
is the anisotropic part of the CR background distribution. It is known
to be about $\alpha\sim10^{-3}$ of the isotropic part $F_{C}$, so
we can estimate $F_{0}\sim\alpha F_{C}$. The last estimate along
with eq.(\ref{eq:FinConstr}) brings us to the maximum beam energy

\begin{equation}
\frac{p_{Bmax}}{mc}\simeq\frac{1}{K}\sqrt{\frac{V_{A}}{c}\frac{A}{\alpha}}\label{eq:Bmax}\end{equation}
where we have introduced the following parameter which is the major
small parameter of the theory

\begin{equation}
K\equiv\frac{c}{l\omega_{c}}=\epsilon\frac{mc}{p}.\label{eq:Kdef}\end{equation}
Here $\omega_{c}$ is the proton cyclotron (non-relativistic) frequency
and $l$ is the maximum turbulence scale beyond which the particles
response becomes adiabatic. Based on the two independent MILAGRO measurements
of the width and the fractional excess of the Beam A, we inferred
earlier the parameter $\epsilon\sim10^{-3}$. Assuming that this value
of $\epsilon$ relates to the $1$TeV median energy of the MILAGRO
collaboration angular analysis, we obtain for $K$ the value $K\sim10^{-6}$.
Taking $V_{A}/c\sim10^{-4}$ and $\alpha\sim A\sim10^{-3}$, we obtain
$p_{Bmax}\sim10$ TeV. This is encouragingly close to the MILAGRO
estimates of the beam cut-off energy. We will consider approaches
to the independent determination of the theory small parameter $K$
and the beam maximum momentum in the next section. Of course, depending
on which of the Milagro beam measurements (the width, excess or cut-off
momentum) is the most reliable, this quantity may be used to determine
$K$ or $l$.

To conclude this section, we estimate the possible losses of the beam
due to the energy dependent curvature and gradient drifts. Assuming
$\nabla\times\mathbf{B}=0$ and a small propagation angle to the magnetic
field (curvature drift dominates), the particle drift velocity can
be written as 

\begin{equation}
\mathbf{V}_{cd}=\frac{p}{mc}\frac{c^{2}}{\omega_{c}B^{2}}\mathbf{B}\times\nabla B\label{eq:drift}\end{equation}
We can estimate particle displacement across the field line upon traveling
a distance of one correlation length $l_{B}$ as $\Delta r\sim r_{g}\left(p\right)l_{B}/R$,
where $R$ is the typical field curvature. The total displacement
from the field line is thus $r\sim r_{g}\sqrt{L_{S}l_{B}}/R\sim r_{g}\sqrt{L_{S}/l}$
where $L_{S}$ is the distance from the source to the observer. The
displacement $r$ may be not much larger than the SNR radius, for
example, so there should be no significant loss of particle flux due
to the drift related spreading.

\section{Distance to the source, beam energy window and the maximum scale
$l$\label{Sec:Distance-to-the}}

Assuming only one free parameter $K=c/l\omega_{c}$ with $l$ being
an (unknown \emph{a priori}) scale of turbulence beyond which particles
are not scattered in pitch angle, we have advanced our theoretical
construction to the point where it successfully matches the three
major MILAGRO observables. These are the angular width of the beam,
its excess and its maximum momentum $p_{Bmax}$. Each of those three
quantities consistently points at the same value of $K\sim10^{-6}$
or $l\sim1pc/B_{\mu G}$. In this section we relate $K$ to the further
two independent quantities. One of these quantities is the maximum
momentum $p_{max}$ of the CRs accelerated in the SNR which may be
responsible for the MILAGRO beam. The other quantity is the distance
to this remnant, $L_{s}$, or to any other source of energetic particles
from which the beam originates. Starting from the source, we represent
the decay of the large scale anisotropic part of the distribution
function as follows (see eqs.{[}\ref{eq:nu},\ref{eq:transp1}{]}
and {[}\ref{eq:PsiExp}{]})

\begin{equation}
F_{S}\left(z,p\right)\sim F_{S}\left(0,p\right)\exp\left[-\mathcal{L}^{-1}\frac{z}{l}\right]\label{eq:Fzdecay}\end{equation}
where $\mathcal{L}^{-1}\left(\epsilon\right)$ is the inverse dimensionless
particle scattering length

\[
\mathcal{L}^{-1}\left(\epsilon\right)=\frac{\lambda_{1}}{6}\left(\delta\ln\frac{1}{\epsilon}+\epsilon^{3/2}\right)\]
and $F_{S}\left(0,p\right)$ is the anisotropic part of the distribution
at the source. As we argued earlier, for the beam to appear at the
Earth ($z=L_{S}$) as observed, $F_{S}\left(L_{S},p\right)$ should
be of the order of the anisotropic part of the local background CRs.
Moreover, $\mathcal{L}^{-1}\left(\epsilon\right)$ has a minimum ($\approx1.7\cdot10^{-3}$)
at $\epsilon=\left(2\delta/3\right)^{2/3}\simeq1.6\cdot10^{-3}$ for
$\delta\equiv V_{A}/c=10^{-4}$. This value of $\epsilon$ is remarkably
close to that inferred earlier from the Milagro measurements of the
beam width and its fractional excess ($\epsilon\simeq2\cdot10^{-3}$).
Since $\epsilon\propto p$, the anisotropic part $F_{S}\left(z,p\right)$
decays rapidly with $p$. Therefore, for the beam to be observable
at $10$ TeV, the distance $L_{s}$ should not significantly exceed
the quantity 

\begin{equation}
L_{Smax}\approx\frac{6l}{\lambda_{1}\epsilon^{3/2}\left(p_{Bmax}\right)}Ln\label{eq:Lmax1}\end{equation}
with 

\[
Ln=\ln\frac{F_{s}\left(0,p\right)}{F_{s}\left(L_{Smax},p\right)}\]
which can be recast as 

\[
L_{Smax}\simeq0.4\frac{c}{\omega_{c}}\left(\frac{mc}{p_{Bmax}}\right)^{3/2}K^{-5/2}Ln\]
or, assuming $K=10^{-6}$, as inferred from the Milagro Spot A parameters,
and $B=3\mu G$, we obtain

\begin{equation}
L_{Smax}\simeq130\cdot Ln\cdot\left(\frac{10TeV}{E_{Bmax}}\right)^{3/2}pc.\label{eq:Lsmax2}\end{equation}
Given that $Ln$ may be a factor of a few, the last estimate constrains
the distance to any SNR, held responsible for the Milagro beam, to
a few hundreds of parsecs. In fact there is also the lower bound to
$L_{S}$ which, being formally a technical one, may still be meaningful.
Indeed, in our calculations of the beam profile, we neglected the
contributions of the eigenfunctions corresponding to the eigenvalues
$\lambda_{n}$ with $n\ge2$. Since $\lambda_{2}\simeq2\lambda_{1}$,
the neglected terms in the spectral expansion of the distribution
function would not contribute near $p\sim p_{Bmax}$, but they could
become essential at lower momenta where $\mathcal{L}^{-1}$ has a
minimum as a function of $p$. That is why we required in Sec.\ref{sec:Particle-propagation}
$\mathcal{L}^{-1}\frac{z}{l}\ga1$. It does not mean, however, that
the beam would not form at these momenta but its shape may change.
Unfortunately, the available Milagro data are not sufficient to distinguish
between the cases of single and multiple beam eigenfunctions. Nevertheless
the apparent absence of a mesoscale anisotropy (\ie scales between
the narrow beam and the first angular harmonics) hints at a relative
unimportance of the higher eigenfunction in the spectral expansion.
If this is the case, then the upper bound on $L_{S}$ given by eq.(\ref{eq:Lsmax2})
should be rather close to the lower bound as well. 

Let us turn to the question of determining the scale $l$. The simplest
possibility is to associate $l$ with the outer scale of the ISM turbulence.
Its typical estimates extend from 1pc (spiral arms) up to 100 pc for
the inter-arm space \citealp{OuterScale08}. However, a 100pc scale
can hardly be relevant to our analysis already for that simple reason
that the Larmor radii of particles of interest are five order of magnitude
smaller. Clearly, such long scales should be attributed to the ambient
field rather than to the particle scattering field component. On the
other hand, as the turbulent energy injected at such long scales cascades
to much shorter scales where the wave can interact with $1-10$ TeV
particles, the spectral energy density is already too low to provide
efficient scattering. Clearly, a realistic estimate of outer scale
of turbulence $l$, relevant for the wave-particle interaction, should
be somewhere between these extremes. If particles are propagating
from an accelerator, there must be energy injection into the GS cascade
at a scale, associated with this accelerator. Obviously, $l$ cannot
exceed the accelerator (shock) radius. It is interesting to note that
the recent optical observations of the SNR 1006 indicate that \emph{ripples
}on the shock surface have a scale \textasciitilde{}1pc \citep{RaymSN1006_07},
which is the preferred scale to match the Milagro data. From the theoretical
standpoint, we need to make an assumption about the accelerator. There
are a few possibilities, such as a nearby SNR or a massive blue star
surrounded by a wind bubble with the termination shock. Each of these,
being magnetically connected with the Earth may accelerate particles
and load the connecting flux rope with both the accelerated particles
and Alfvenic turbulence. In order to avoid further uncertainties associated
with the accelerator, we assume that the turbulence is driven primarily
by escaping particles. This is almost certainly the case, once particles
escape at the rate sufficient to be detected at the Earth. The turbulence,
however may significantly decay along the flux rope due to the relaxation
of initially strongly unstable (anisotropic) particle distribution
and due to lateral losses of particles and waves. Note that if these
are significant, one should replace $zD\left(\mu\right)\to\int Ddz$
in our treatment of particle propagation in Sec.\ref{sec:Particle-propagation}.

The mechanisms of particle escape from a SNR shock, for example, are
many \citep{DruryNeutral96,mdj02,MD06,BlasiEscape09,RevilleEsc09}.
In almost all cases the escaping particles are close to the maximum
energy achievable in the accelerator and have an anisotropic momentum
distribution. Therefore they should drive Alfven waves at a scale
$l\sim r_{g}\left(p_{max}\right)$. Since $r_{g}\left(p\right)\simeq10^{-6}B_{\mu G}^{-1}\left(p/mc\right)$pc,
to recover the scale $l\sim1$ pc, inferred earlier from the beam
parameters, it is necessary to assume $E_{max}\sim3$ PeV (for $B_{\mu G}\sim3$)
or precisely the 'knee' energy.

\section{Summary and discussion\label{sec:Summary-and-discussion}}

The principal results of this paper are as follows. Assuming only
a \emph{large scale} anisotropic distribution of CRs (generated, for
example by a nearby accelerator, such as a SNR) and a \citet{Goldr95}
(GS) cascade of Alfvenic turbulence originating from some scale $l$,
which is the longest scale relevant for the wave-particle interactions,
we calculated the propagation of the CRs down their gradient along
the interstellar magnetic field. It is found that the CR distribution
develops a characteristic angular shape consisting of a large scale
anisotropic part (first eigenfunction of the pitch-angle scattering
operator) superposed by a beam, tightly focused in the momentum space
in the local field direction. The large scale anisotropy carries the
\emph{momentum dependence }of the source,\emph{ }while both the beam
angular width and its fractional excess (with respect to the large
scale anisotropic component) grow with momentum (as $\sqrt{p}$ and
$p$, respectively). Apart from the width and the fractional excess
of the beam, the theory predicts its maximum momentum on the ground
that beyond this momentum the beam destroys itself. All the three
quantities are completely determined by the turbulence scale $l$.
Even if $l$ is considered unknown, it can be inferred from any of
the three independent MILAGRO measurements. These are the width, the
fractional excess and the maximum energy of the beam, and all the
three consistently imply the same scale $l\sim$1 pc. The calculated
beam maximum momentum encouragingly agrees with that measured by MILAGRO
(\textasciitilde{}10 TeV/c). The theoretical value for the angular
width of the beam is found to be $\Delta\vartheta\simeq4\sqrt{\epsilon}$,
where $\epsilon=r_{g}\left(p\right)/l\ll1$. The beam fractional excess
related to the large scale anisotropic part of the CR distribution
is $\simeq50\epsilon$. Both quantities also match the Milagro results
for $E\sim1-2$ TeV. So the beam has a momentum scaling that is one
power shallower than the CR carrier, it is drawn from. This finding
will receive a due discussion. 

Obviously, the determination of the absolute value of the beam excess
would require the source intensity. For the lack of such information,
an indirect inference was made in Sec.\ref{sub:Observational-consequences}
about the galactic CR (GCR) large scale anisotropy being of the same
order as the large scale anisotropy responsible for the beam.

Below the rationale for this admission is given which we open with
the following notations:
\begin{itemize}
\item $\tilde{F}_{GCR}$ and $\bar{F}_{GCR}$ are the large scale anisotropic
and isotropic parts of the galactic (not assumed to be related to
the source of the beam) distribution, respectively
\item $\tilde{F}_{S}$ and $\bar{F}_{S}$ (i.e. $f_{\infty}$ in sec.\ref{sec:Particle-propagation})
are the similar quantities related to the source of the beam at the
distance $L_{S}$
\item $F_{B}$ is the beam distribution on the top of $\tilde{F}_{S}$
\end{itemize}
Unless the source of the beam is also responsible for the GCR, the
quantities $\tilde{F}_{S}$ and $\tilde{F}_{GCR}$ are independent
of each other and cannot be related since the source intensity, the
distance to it, $L_{S}$ and the losses are unknown. Even if the beam
and its carrier $\tilde{F}_{S}$ propagate without significant losses
(or suffer similar losses), the current theory determines only a fractional
excess $F_{B}/\tilde{F}_{S}\sim50\epsilon$ (independent of $L_{S}$).
For the same reasons we do not know what is the source contribution
$\bar{F}_{S}$ into the total isotropic CR background $\bar{F}{}_{GCR}+\bar{F}_{S}$.
However, since the beam A is not observed at the minimum%
\footnote{It is interesting to note that \citet{Milagro08PRL} point out that
there is a deep deficit bordering the excess regions. This deficit
could be identified as a minimum of the dominant eigenfunction, but
they attribute it to the effect of including the excess regions into
the background. In other words the deficit is an artefact of the data
analysis.%
} of the (measured) total large scale $\tilde{F}_{S}+\tilde{F}_{GCR}$
as it would, were $\tilde{F}_{S}\gg\tilde{F}_{GCR}$ the case (see
Fig.\ref{fig:Eigenfunction}), we infer $\tilde{F}_{GCR}\ga\tilde{F}_{S}$.
Furthermore, since $F_{B}/\tilde{F}_{S}$ is calculated and $\tilde{F}_{S}+\tilde{F}_{GCR}$
is measured along with $F_{B}$, the both quantities $\tilde{F}_{S}$
and $\tilde{F}_{GCR}$ can also be determined. 

We found that $F_{B}/\tilde{F}_{S}\sim0.1$ for $l\sim1$pc which
was, in turn, deduced from two other independent measurements (beam
width $\Delta\vartheta$ and its maximum energy $E_{Bmax}$). Since
Milagro measurements indicate that $F_{B}/\left(\tilde{F}_{GCR}+\tilde{F}_{S}\right)\sim0.1$,
we conclude that $\tilde{F}_{S}\sim\tilde{F}_{GCR}$. A more specific
relation between the two would not be meaningful since the measurements
of $F_{B}\left(p\right)$ are rather limited. If particle losses from
the flux tube are negligible, it follows that $\tilde{F}_{S}\left(L_{S}\right)\sim\tilde{F}_{S}\left(0\right)$
for $p\ll p_{Bmax}$ ($0$- being the source position). 

These findings allow us to speculate about the possible source of
the beam. First, if the source is an active accelerator that emits
strongly anisotropic particle flux, the last relation implies that
$\bar{F}_{S}\sim\tilde{F}_{S}$. Since by observations $\tilde{F}_{S}\ll\bar{F}_{GCR}$,
such source cannot contribute significantly to the 'knee' region at
$\simeq3$ PeV. In this case our inferences of $l$ from three independent
measurements --all strikingly pointing at the $3$ PeV accelerator
cut-off energy (with $l\sim r_{g}\left(E_{max}\right)$)-- must be
either a coincidence or a different mechanism couples the galactic
'knee' particles with the scale of the turbulence that generates the
beam. If it is not a coincidence and the source contributes significantly
to the observed CR background, the escaping particle flux should be
quasi-isotropic, $\tilde{F}_{S}\ll\bar{F}_{S}$ (to allow for $\bar{F}_{S}\sim\bar{F}_{GCR}$).
In combination with the assumption that particles escape in the wide
range 1TeV -3 PeV (to both form the beam and to inject MHD energy
at the scale $l$), the source is unlikely to be an active accelerator,
but rather a region of an enhanced CR density, with a steep cut-off
at $\simeq3$ PeV. The near isotropy at the source is not inconsistent
with a currently working accelerator, but escape in such a broad energy
range probably is. Indeed, at least the available (known to us) mechanisms,
that offer a broad energy escape from a SNR along with the spectrum
steepening (\ie spectral break, starting 1-2 orders of magnitude
below the cut-off, \eg \citealp{MDS05,MD06}) seem to fall short
to cover three orders of magnitude in energy. Moreover for the source
to be a recent accelerator (such as a recent SNR, suggested by \citealp{ErlykinWolfendale97}
with the spectrum $E^{-2}$) the mechanism should be found that makes
the spectrum of the escaping particles at least $0.5$ steeper (and
still steeper if the acceleration was strongly nonlinear). Combined
with the nonlinear acceleration (which is required in \citealp{MD06}
model) this would make an acceptable spectrum but again, it is not
clear how these particles can initiate the MHD cascade at such a long
scale, to ensure the required value of $l$. 

Our argument against the beam and the bulk CR $\tilde{F}_{GCR}$ coming
from the same source is that the observed beam is not located at the
minimum of the angular distribution of the first eigenfunction, so
we need to allow for a second component. This is largely a technical
limitation, stemming from 1-D transport model, in which $F_{B}$ and
$\tilde{F}_{S}$ are coupled, as wells as from the single eigenfunction
approximation. By removing this latter simplification alone (which
is probably even necessary for an accurate description of $\tilde{F}_{S}$
at TeV- energies, Sec.\ref{Sec:Distance-to-the}) the above constraint
can be relaxed. Another possibility is a lateral diffusion and drifts
of $\tilde{F}_{S}$-component from the flux tube. 

An interesting obvious conjecture from the common origin of the beam
and $\tilde{F}_{GCR}$ would be that the proton 'knee' at $\simeq3$
PeV is also of the same origin as the beam. However, the beam spectrum
is calculated to be one power flatter than its carrier. According
to Milagro the beam index is about $1.5$, so that the carrier should
have an index $\simeq2.5$ which is closer to the GCR than to a hypothetical
'recent SNR'. In particular, this would not support the single source
hypothesis of the GCR 'knee' \citet{ErlykinWolfendale97}. Equally
problematic would be an active accelerator scenario, unless the steepening
mechanisms of the run-away CR mentioned above can be adopted after
due modifications.

All told, the beam is likely to be at least partly drawn from the
GCR (due to the relation between the indices $q_{B}\simeq q_{GCR}-1$)
but the GS- turbulence that creates the beam must be driven by considerably
more energetic, $\ga1$ PeV particles (due to the constraint, $l\simeq r_{g}$),
unless the spiral-arm 1pc value \citet{OuterScale08} for $l$ is,
indeed, acceptable. To explore the possibility of the GCR origin of
the beam, an extension of the above model is necessary. At a minimum,
the model should include the transport of energetic particles across
the flux tube. On the other hand, the particle beaming processes should
remain similar to that described in Sec.\ref{sec:Pitch-angle-scattering-of}.
However, such consideration is out of the scope of this paper, particularly
because the transport across the flux tube requires a separate study. 

Yet another possibility is that the required GS cascade starts in
the local interstellar cloud (LIC). It has a suitable size of $\sim5$
pc \citep{LIC00} and there would be no problem with the spectrum
slope since the beam would be drawn from the GCR with the 'right'
spectral index $q_{GCR}\simeq2.7$. Whether the turbulence energy
can be injected at the required scale remains to be studied. If it
can, the above transport and beam focusing mechanism would be applicable
since a parsec wave length and the GS cascade are the only requirements
to draw the beam out of the background CR distribution.

To conclude, the model presented in this paper offers an explanation
of the most pronounced Milagro beam A, while there are two more. One
of them is the beam B, $\sim50^{\circ}$ away from beam A and the
second one is in the Cygnus loop area $\sim100^{\circ}$ away. Any
attempt to incorporate those two beams into our current model would
be speculative. We merely note that the local ISM environment is complicated
indeed thus offering many possibilities in explaining various CR anomalies
(\eg \citealp{Amenomori07}). Approaches to the explanation of all
three beams based on such a complexity, could hardly pass the Occam's
razor test. In contrast, the model suggested in this paper is devoid
of free parameters, if the knee energy at $\sim3$PeV can be associated
with the maximum CR energy of the source of the Beam A. Even though
such an association is not proven, our propagation model predicts
the following three beam characteristics: its width, fractional excess
and maximum energy to be the functions of a single quantity, the longest
wave-particle interaction scale $l$. They all give the correct MILAGRO
values for $l\simeq1$ pc, which is unlikely to be coincidental. However,
the exact origin of this particular value remains unclear.

\acknowledgements{}

LD and MM acknowledge the hospitality of KITP in Santa Barbara during
the Program Particle Acceleration in Astrophysical Plasmas, July 26-October
3, 2009. The work of PD and MM is supported by NASA under the Grants
NNX 07AG83G and NNX09AT94G as and by the Department of Energy, Grant
No. DE-FG02-04ER54738.

\appendix{}

\section{Appendix\label{sec:AppendixA}}

In this Appendix we provide a sketch of the derivation of the pitch-angle
diffusion coefficient $D_{\mu\mu}$ for the anisotropic turbulence
of Alfven waves suggested by \citet{Goldr95}. We follow a standard
line of argument (e.g., \citealp{Voelk73}). However, we include the
finite auto-correlation time as required by the GS spectrum. We start
with the equation for the particle momentum $\mathbf{p}$:

\begin{equation}
\frac{d\mathbf{p}}{dt}=\Omega\mathbf{p}\times\mathbf{B}/B_{0}\label{eq:ApAeqm}\end{equation}
where $\Omega=eB_{0}/p$, $p\gg mc$ and $B_{0}$ is the magnitude
of the unperturbed magnetic field $\mathbf{B}_{0}$, assumed to be
in $z$-direction. We also decompose the total magnetic field $\mathbf{B}$
in the following standard way

\begin{equation}
\mathbf{B}=B_{0}\hat{z}+\sum_{\mathbf{k}}\mathbf{B_{\mathbf{k}}}e^{i\mathbf{kr}}\label{eq:ABdecomp}\end{equation}
where $\hat{z}$ is the unit vector along $z$-axis. Note that for
the shear Alfven waves, $\mathbf{B_{k}}\perp\mathbf{k},\hat{\, z}$.
As usual, we introduce a spherical coordinate system in the momentum
space with the axis along the unperturbed magnetic field: $p_{\parallel}=p\mu=\mathbf{p}\cdot\hat{z}$,
$p_{\perp}=p\sqrt{1-\mu^{2}}$, $p_{x}+ip_{y}=p_{\perp}\exp\left(i\phi\right)$.
The corresponding notations in $\mathbf{k}$-space are $k_{\parallel}=\mathbf{k}\cdot\hat{z}$,
$k_{x}+ik_{y}=k_{\perp}\exp\left(\alpha_{\mathbf{k}}\right),$ and
similarly for $\mathbf{B_{k}}$: $B_{\mathbf{k},x}+iB_{\mathbf{k},y}=B_{\mathbf{k}}\exp\left(i\chi_{\mathbf{k}}\right)$,
where $\chi_{\mathbf{k}}=\alpha_{\mathbf{k}}\pm\pi/2$, where the
$\pm$ corresponds to the direction of the wave propagation, $\omega=\pm\left|k_{\parallel}\right|V_{A}$.
With this notations, also using the relation 

\[
\mathbf{kr}=k_{\parallel}v_{\parallel}t-\xi\sin\left(\phi-\alpha_{\mathbf{k}}\right),\]
with $\xi=k_{\perp}v_{\perp}/\Omega$, from eq.(\ref{eq:ApAeqm})
we obtain

\begin{equation}
\frac{d\mu}{dt}=\pm\frac{\Omega}{B_{0}}\sqrt{1-\mu^{2}}\sum_{\mathbf{k},n}B_{\mathbf{k}}e^{ik_{\parallel}v_{\parallel}t+in\left(\Omega t-\phi_{0}+\alpha_{\mathbf{k}}\right)}\frac{n}{\xi}J_{n}\left(\xi\right)\label{eq:Admudt}\end{equation}
where $\phi_{0}$ comes from the unperturbed particle orbit $\phi=\phi_{0}-\Omega t$
and $J_{n}$ stands for the Bessel function. Denoting by $\Delta\mu$
the variation of $\mu$ in time $t$, for an ensemble averaged $\left\langle \Delta\mu^{2}\right\rangle $
we obtain

\[
\left\langle \Delta\mu^{2}\right\rangle =\Omega^{2}\left(1-\mu^{2}\right)\sum_{\mathbf{k},n}\iintop_{0}^{t}dt^{\prime}dt^{\prime\prime}I\left(k_{\parallel},k_{\perp},t^{\prime}-t^{\prime\prime}\right)e^{\left(ik_{\parallel}v_{\parallel}+in\Omega\right)\left(t^{\prime}-t^{\prime\prime}\right)}\frac{n^{2}}{\xi^{2}}J_{n}^{2}\left(\xi\right)\]
where 

\[
I_{\mathbf{k}}\left(t^{\prime}-t^{\prime\prime}\right)=\left\langle B_{\mathbf{k}}\left(t^{\prime}\right)\bar{B}_{\mathbf{k}}\left(t^{\prime\prime}\right)\right\rangle /B_{0}^{2}\]
is assumed to be axially symmetric in $\mathbf{k}-$space. Extracting
the secular term from the last equation, we obtain eq.(\ref{eq:Dmm1}).
Note that it can be further simplified by performing the summation
in $n$

\[
D_{\mu\mu}=-\left(1-\mu^{2}\right)\sum_{\mathbf{k}}\frac{1}{\xi^{2}}\intop_{0}^{\infty}I\left(k_{\parallel},k_{\perp},\tau\right)e^{ik_{\parallel}v_{\parallel}\tau}d\tau\frac{\partial^{2}}{\partial\tau^{2}}J_{0}\left(2\xi\sin\frac{\Omega\tau}{2}\right).\]

\pagebreak

\begin{thebibliography}{42}
\expandafter\ifx\csname natexlab\endcsname\relax\def\natexlab#1{#1}\fi

\bibitem[{{Abdo} {et~al.}(2008){Abdo}, {Allen}, {Aune}, {Berley}, {Blaufuss},
  {Casanova}, {Chen}, {Dingus}, {Ellsworth}, {Fleysher}, {Fleysher},
  {Gonzalez}, {Goodman}, {Hoffman}, {H{\"u}ntemeyer}, {Kolterman}, {Lansdell},
  {Linnemann}, {McEnery}, {Mincer}, {Nemethy}, {Noyes}, {Pretz}, {Ryan},
  {Parkinson}, {Shoup}, {Sinnis}, {Smith}, {Sullivan}, {Vasileiou}, {Walker},
  {Williams}, \& {Yodh}}]{Milagro08PRL}
{Abdo}, A.~A., {Allen}, B., {Aune}, T., {Berley}, D., {Blaufuss}, E.,
  {Casanova}, S., {Chen}, C., {Dingus}, B.~L., {Ellsworth}, R.~W., {Fleysher},
  L., {Fleysher}, R., {Gonzalez}, M.~M., {Goodman}, J.~A., {Hoffman}, C.~M.,
  {H{\"u}ntemeyer}, P.~H., {Kolterman}, B.~E., {Lansdell}, C.~P., {Linnemann},
  J.~T., {McEnery}, J.~E., {Mincer}, A.~I., {Nemethy}, P., {Noyes}, D.,
  {Pretz}, J., {Ryan}, J.~M., {Parkinson}, P.~M.~S., {Shoup}, A., {Sinnis}, G.,
  {Smith}, A.~J., {Sullivan}, G.~W., {Vasileiou}, V., {Walker}, G.~P.,
  {Williams}, D.~A., \& {Yodh}, G.~B. 2008, Physical Review Letters, 101,
  221101

\bibitem[{{Achterberg}(1983)}]{Acht83FireHose}
{Achterberg}, A. 1983, \aap, 119, 274

\bibitem[{{Amenomori} {et~al.}(2007){Amenomori}, {Ayabe}, {Bi}, {Chen}, {Cui},
  {Danzengluobu}, {Ding}, {Ding}, {Feng}, {Feng}, {Feng}, {Gao}, {Geng}, {Guo},
  {He}, {He}, {Hibino}, {Hotta}, {Hu}, {Hu}, {Huang}, {Huang}, {Jia}, {Kajino},
  {Kasahara}, {Katayose}, {Kato}, {Kawata}, {Labaciren}, {Le}, {Li}, {Li},
  {Lou}, {Lu}, {Lu}, {Meng}, {Mizutani}, {Mu}, {Munakata}, {Nagai}, {Nanjo},
  {Nishizawa}, {Ohnishi}, {Ohta}, {Onuma}, {Ouchi}, {Ozawa}, {Ren}, {Saito},
  {Saito}, {Sakata}, {Sako}, {Sasaki}, {Shibata}, {Shiomi}, {Shirai},
  {Sugimoto}, {Takita}, {Tan}, {Tateyama}, {Torii}, {Tsuchiya}, {Udo}, {Wang},
  {Wang}, {Wang}, {Wang}, {Wu}, {Xue}, {Yamamoto}, {Yan}, {Yang}, {Yasue},
  {Ye}, {Yu}, {Yuan}, {Yuda}, {Zhang}, {Zhang}, {Zhang}, {Zhang}, {Zhang},
  {Zhang}, {Zhaxisangzhu}, \& {Zhou}}]{Amenomori07}
{Amenomori}, M., {Ayabe}, S., {Bi}, X.~J., {Chen}, D., {Cui}, S.~W.,
  {Danzengluobu}, {Ding}, L.~K., {Ding}, X.~H., {Feng}, C.~F., {Feng}, Z.,
  {Feng}, Z.~Y., {Gao}, X.~Y., {Geng}, Q.~X., {Guo}, H.~W., {He}, H.~H., {He},
  M., {Hibino}, K., {Hotta}, N., {Hu}, H., {Hu}, H.~B., {Huang}, J., {Huang},
  Q., {Jia}, H.~Y., {Kajino}, F., {Kasahara}, K., {Katayose}, Y., {Kato}, C.,
  {Kawata}, K., {Labaciren}, {Le}, G.~M., {Li}, A.~F., {Li}, J.~Y., {Lou}, Y.,
  {Lu}, H., {Lu}, S.~L., {Meng}, X.~R., {Mizutani}, K., {Mu}, J., {Munakata},
  K., {Nagai}, A., {Nanjo}, H., {Nishizawa}, M., {Ohnishi}, M., {Ohta}, I.,
  {Onuma}, H., {Ouchi}, T., {Ozawa}, S., {Ren}, J.~R., {Saito}, T., {Saito},
  T.~Y., {Sakata}, M., {Sako}, T.~K., {Sasaki}, T., {Shibata}, M., {Shiomi},
  A., {Shirai}, T., {Sugimoto}, H., {Takita}, M., {Tan}, Y.~H., {Tateyama}, N.,
  {Torii}, S., {Tsuchiya}, H., {Udo}, S., {Wang}, B., {Wang}, H., {Wang}, X.,
  {Wang}, Y.~G., {Wu}, H.~R., {Xue}, L., {Yamamoto}, Y., {Yan}, C.~T., {Yang},
  X.~C., {Yasue}, S., {Ye}, Z.~H., {Yu}, G.~C., {Yuan}, A.~F., {Yuda}, T.,
  {Zhang}, H.~M., {Zhang}, J.~L., {Zhang}, N.~J., {Zhang}, X.~Y., {Zhang}, Y.,
  {Zhang}, Y., {Zhaxisangzhu}, \& {Zhou}, X.~X. 2007, 932, 283

\bibitem[{{Bell}(2004)}]{Bell04}
{Bell}, A.~R. 2004, \mnras, 353, 550

\bibitem[{{Beresnyak} \& {Lazarian}(2009)}]{BereznLaz09}
{Beresnyak}, A., \& {Lazarian}, A. 2009, \apj, 702, 1190

\bibitem[{{Beresnyak} {et~al.}(2010){Beresnyak}, {Yan}, \&
  {Lazarian}}]{BereznLazar10}
{Beresnyak}, A., {Yan}, H., \& {Lazarian}, A. 2010, ArXiv e-prints

\bibitem[{{Caprioli} {et~al.}(2009){Caprioli}, {Blasi}, \&
  {Amato}}]{BlasiEscape09}
{Caprioli}, D., {Blasi}, P., \& {Amato}, E. 2009, \mnras, 396, 2065

\bibitem[{{Chandran}(2000)}]{ChandranGS00PhRvL}
{Chandran}, B.~D.~G. 2000, Physical Review Letters, 85, 4656

\bibitem[{{Cho} \& {Vishniac}(2000)}]{ChoVishn00}
{Cho}, J., \& {Vishniac}, E.~T. 2000, \apj, 539, 273

\bibitem[{{Drury} \& {Aharonian}(2008)}]{DruryAharMILAGRO08}
{Drury}, L.~O.~C., \& {Aharonian}, F.~A. 2008, Astroparticle Physics, 29, 420

\bibitem[{{Drury} {et~al.}(1996){Drury}, {Duffy}, \& {Kirk}}]{DruryNeutral96}
{Drury}, L.~O.~C., {Duffy}, P., \& {Kirk}, J.~G. 1996, \aap, 309, 1002

\bibitem[{{Drury} \& {Falle}(1986)}]{DruryFal86}
{Drury}, L.~O.~C., \& {Falle}, S.~A.~E.~G. 1986, \mnras, 223, 353

\bibitem[{{Ellison} {et~al.}(1996){Ellison}, {Baring}, \&
  {Jones}}]{Ellison1996}
{Ellison}, D.~C., {Baring}, M.~G., \& {Jones}, F.~C. 1996, \apj, 473, 1029

\bibitem[{{Erlykin} \& {Wolfendale}(1997)}]{ErlykinWolfendale97}
{Erlykin}, A.~D., \& {Wolfendale}, A.~W. 1997, Journal of Physics G Nuclear
  Physics, 23, 979

\bibitem[{{Goldreich} \& {Sridhar}(1995)}]{Goldr95}
{Goldreich}, P., \& {Sridhar}, S. 1995, \apj, 438, 763

\bibitem[{{Goldreich} \& {Sridhar}(1997)}]{goldr97}
---. 1997, \apj, 485, 680

\bibitem[{{Gurevich}(1961)}]{Gurevich61RunAway}
{Gurevich}, A.~V. 1961, Soviet Journal of Experimental and Theoretical Physics,
  12, 904

\bibitem[{{Haverkorn} {et~al.}(2008){Haverkorn}, {Brown}, {Gaensler}, \&
  {McClure-Griffiths}}]{OuterScale08}
{Haverkorn}, M., {Brown}, J.~C., {Gaensler}, B.~M., \& {McClure-Griffiths},
  N.~M. 2008, \apj, 680, 362

\bibitem[{{Jokipii}(1966)}]{Jokipii66}
{Jokipii}, J.~R. 1966, \apj, 146, 480

\bibitem[{{Kang} {et~al.}(1992){Kang}, {Jones}, \& {Ryu}}]{KangJR92}
{Kang}, H., {Jones}, T.~W., \& {Ryu}, D. 1992, \apj, 385, 193

\bibitem[{{Kennel} \& {Engelmann}(1966)}]{Kennel66}
{Kennel}, C.~F., \& {Engelmann}, F. 1966, Physics of Fluids, 9, 2377

\bibitem[{{Kirk} \& {Duffy}(1999)}]{KirkDuffyRelS99}
{Kirk}, J.~G., \& {Duffy}, P. 1999, Journal of Physics G Nuclear Physics, 25,
  163

\bibitem[{{Kirk} \& {Schneider}(1987)}]{KirkSchn87}
{Kirk}, J.~G., \& {Schneider}, P. 1987, \apj, 315, 425

\bibitem[{{Kruskal} \& {Bernstein}(1964)}]{KruskalBernstein64}
{Kruskal}, M.~D., \& {Bernstein}, I.~B. 1964, Physics of Fluids, 7, 407

\bibitem[{{Malkov} \& {Diamond}(2006)}]{MD06}
{Malkov}, M.~A., \& {Diamond}, P.~H. 2006, \apj, 642, 244

\bibitem[{{Malkov} {et~al.}(2002){Malkov}, {Diamond}, \& {Jones}}]{mdj02}
{Malkov}, M.~A., {Diamond}, P.~H., \& {Jones}, T.~W. 2002, \apj, 571, 856

\bibitem[{{Malkov} {et~al.}(2005){Malkov}, {Diamond}, \& {Sagdeev}}]{MDS05}
{Malkov}, M.~A., {Diamond}, P.~H., \& {Sagdeev}, R.~Z. 2005, \apjl, 624, L37

\bibitem[{{Malkov} \& {Voelk}(1995)}]{mv95}
{Malkov}, M.~A., \& {Voelk}, H.~J. 1995, \aap, 300, 605

\bibitem[{{Maron} \& {Goldreich}(2001)}]{MaronGoldr01}
{Maron}, J., \& {Goldreich}, P. 2001, \apj, 554, 1175

\bibitem[{{Raymond} {et~al.}(2007){Raymond}, {Korreck}, {Sedlacek}, {Blair},
  {Ghavamian}, \& {Sankrit}}]{RaymSN1006_07}
{Raymond}, J.~C., {Korreck}, K.~E., {Sedlacek}, Q.~C., {Blair}, W.~P.,
  {Ghavamian}, P., \& {Sankrit}, R. 2007, \apj, 659, 1257

\bibitem[{{Redfield} \& {Linsky}(2000)}]{LIC00}
{Redfield}, S., \& {Linsky}, J.~L. 2000, \apj, 534, 825

\bibitem[{{Reville} {et~al.}(2009){Reville}, {Kirk}, \& {Duffy}}]{RevilleEsc09}
{Reville}, B., {Kirk}, J.~G., \& {Duffy}, P. 2009, \apj, 694, 951

\bibitem[{Richardson(1918)}]{Richardson1918}
Richardson, R. G.~D. 1918, American Journal of Mathematics, 40, 283

\bibitem[{{Rowlands} {et~al.}(1966){Rowlands}, {Shapiro}, \&
  {Shevchenko}}]{RowlandsShapShev66}
{Rowlands}, J., {Shapiro}, V.~D., \& {Shevchenko}, V.~I. 1966, Soviet Journal
  of Experimental and Theoretical Physics, 23, 651

\bibitem[{Sagdeev \& Shafranov(1961)}]{SagdShafr61}
Sagdeev, R.~Z., \& Shafranov, V.~D. 1961, {Soviet Phys. JETP}, 12, 130

\bibitem[{{Salvati} \& {Sacco}(2008)}]{SalvatiMilagro08}
{Salvati}, M., \& {Sacco}, B. 2008, \aap, 485, 527

\bibitem[{{Shapiro} {et~al.}(1998){Shapiro}, {Quest}, \&
  {Okolicsanyi}}]{ShapiroQuest98GeoRL}
{Shapiro}, V.~D., {Quest}, K.~B., \& {Okolicsanyi}, M. 1998, \grl, 25, 845

\bibitem[{{Sridhar} \& {Goldreich}(1994)}]{SridhGoldr94}
{Sridhar}, S., \& {Goldreich}, P. 1994, \apj, 432, 612

\bibitem[{Vedenov {et~al.}(1962)Vedenov, Velikhov, \& Sagdeev}]{VVSQL62}
Vedenov, A.~A., Velikhov, E.~P., \& Sagdeev, R.~Z. 1962, NUCLEAR FUSION, 465

\bibitem[{{V{\"o}lk}(1973)}]{Voelk73}
{V{\"o}lk}, H.~J. 1973, \apss, 25, 471

\bibitem[{{Yan} \& {Lazarian}(2002)}]{YanLazar02}
{Yan}, H., \& {Lazarian}, A. 2002, Physical Review Letters, 89, B1102+

\bibitem[{{Zank} {et~al.}(1990){Zank}, {Axford}, \& {McKenzie}}]{ZankAM90}
{Zank}, G.~P., {Axford}, W.~I., \& {McKenzie}, J.~F. 1990, \aap, 233, 275

\end{thebibliography}
\end{document}